\documentclass[11pt]{article}
\usepackage[top=1in, bottom=1in, right=1in, left=1in]{geometry} 
\usepackage{epsfig,fullpage,multirow,amsmath,amsfonts,natbib,latexsym, bm, parskip, setspace, arydshln ,titlesec}
\usepackage[affil-it]{authblk}
\usepackage{arydshln}
\usepackage[table]{xcolor}

\titlespacing*{\section}
{0pt}{0pt}{0pt}
\titlespacing*{\subsection}
{0pt}{0pt}{0pt}
\titlespacing*{\subsubsection}
{0pt}{0pt}{0pt}

\newcommand{\bfone}{{\bf 1}}

\newcommand{\bfbeta}{{\bm \beta}}
\newcommand{\bfalpha}{{\bm \alpha}}
\newcommand{\bftheta}{{\bm \theta}}
\newcommand{\bfmu}{{\bm \mu}}

\newcommand{\bfzeta}{{\bm \zeta}}
\newcommand{\bfomega}{{\bm \omega}}
\newcommand{\bfnu}{{\bm \nu}}

\newcommand{\bfepsilon}{{\bm \epsilon}}
\newcommand{\bfxi}{{\bm \xi}}
\newcommand{\bfSigma}{{\bm \Sigma}}
\newcommand{\bfsigma}{{\bm \sigma}}

\newcommand{\bfLambda}{{\bm \Lambda}}
\newcommand{\bfPsi}{{\bm \Psi}}

\def\bfm{\mathbf{m}}
\def\bfy{\mathbf{y}}
\def\bfx{\mathbf{x}}

\def\bfZ{\mathbf{Z}}

\def\bfV{\mathbf{V}}
\def\bfs{\mathbf{s}}

\def\bfB{\mathbf{B}}

\def\bfX{\mathbf{X}}
\def\bfY{\mathbf{Y}}
\def\bfA{\mathbf{A}}

\def\bfU{\mathbf{U}}

\def\bfI{\mathbf{I}}

\def\bfW{\mathbf{W}}
\def\bfM{\mathbf{M}}

\def\bfu{\mathbf{u}}
\def\neg{\mbox{-}}
\def\inv{^{-1}}
\def\bfzero{\mathbf{0}}
\def\smneg{\mbox{-}}
\def\bfzero{\mathbf{0}}
\def\neg{\mbox{-}}
\def\inv{^{-1}}
\def\cur{^{[t]}}
\def\prev{^{[t-1]}}
\def\smneg{\mbox{-}}

\makeatletter
\def\@maketitle{%
  \newpage
  \null
  \vskip 2em%
  \begin{center}%
  \let \footnote \thanks
    {\Large\bfseries \@title \par}%
    \vskip 1em%
    {\normalsize
      \lineskip .2em%
      \begin{tabular}[t]{c}%
        \@author
      \end{tabular}\par}%

    {\small \@date}%
  \end{center}%
 \vskip 1em
  \par
}

\title{Bayesian Spatial Binary Classification}

 \author[1]{Candace Berrett\thanks{cberrett@stat.byu.edu}}\affil[1]{Department of Statistics, Brigham Young University, Provo, UT, USA}
 
 \author[2]{Catherine A.\ Calder\thanks{calder@stat.osu.edu}}\affil[2]{Department of Statistics, The Ohio State University, Columbus, OH, USA}

\date{\today}

\begin{document}

\maketitle

\begin{abstract}
In analyses of spatially-referenced data, researchers often have one of two goals: to quantify relationships between a response variable and covariates while accounting for residual spatial dependence or to predict the value of a response variable at unobserved locations.  In this second case, when the response variable is categorical, prediction can be viewed as a classification problem.  Many classification methods either ignore response-variable/covariate relationships and rely only on spatially proximate observations for classification, or they ignore spatial dependence and use only the covariates for classification.  The Bayesian spatial generalized linear (mixed) model offers a tool to accommodate both spatial and covariate sources of information in classification problems.  In this paper, we formally define spatial classification rules based on these models.  We also take a close look at two of these models that have been proposed in the literature, namely the probit versions of the spatial generalized linear model (SGLM) and the Bayesian spatial generalized linear mixed model (SGLMM).  We describe the implications of the seemingly slight differences between these models for spatial classification and explore the issue of robustness to model misspecification through a simulation study.  We also provide an overview of alternatives to the SGLM/SGLMM-based classifiers and illustrate the various methods using satellite-derived land cover data from Southeast Asia.
\end{abstract}

\section{Introduction}
Prediction of unobserved binary or categorical variables can be cast as a {classification} problem, where a classification rule is used to assign an unobserved variable to a class, or category, based on a collection of observed \emph{inputs} (e.g., predictors or covariate information).  A classification rule is determined by a \emph{decision function}, or a function of the inputs, which can be derived from either an underlying statistical model (e.g., logistic regression and discriminant analysis) or an algorithmic method such as support vector machines (SVM) and k-nearest neighbors (kNN) (see \citealp{hast:etal:2001}, for an overview).  In this paper, we consider the spatial classification problem.  That is, we seek to define classification rules to assign an unobserved variable associated with a spatial location (a particular point in a continuously-indexed spatial domain or an area in a discretely-indexed spatial domain) to one or more discrete classes.  We refer to this spatial location as the \textit{focal location} and the area surrounding it as the \textit{neighborhood} of the focal location.   

In classification problems involving spatially-referenced variables, often neighboring values of the unobserved/unknown variable should be used as inputs to the decision function, along with other inputs associated with the focal location and its neighbors.  For example, binary or categorical images derived from satellite remote sensing often contain unobserved locations (or, ``pixels'' in this setting) due to errors in processing the raw data or measurement complications such as cloud cover.   In these situations, formal classification methods are needed to assign values to the unobserved location so that the images can be used for various purposes in scientific investigations.  While values of inputs associated with the focal location (e.g., land cover) may contain valuable information, knowledge of the class of neighboring locations may also be useful in classifying the focal location correctly.  As we will illustrate, classification rules that rely on neighboring observations can be derived from the Bayesian spatial generalized linear and generalized linear mixed models (SGLMs and SGLMMs, respectively).  

Some spatial classification methods have been proposed in the literature, many of which were motivated by remote sensing applications where the measured spectra serve as inputs/covariates.  Unlike the classifiers derived from SGLM/SGLMMs in which spatial proximity explains the patterning of a categorical outcome after accounting for covariates (i.e., spatial dependence is in the `residuals'), these alternative spatial classifiers take advantage of spatial dependence in the covariates associated with each location.  For example, \cite{swit:1980} and \cite{mard:1984} build on the traditional linear discriminant analysis by augmenting the covariates associated with the focal location with the covariates of neighboring locations in determining classification rules.  Building on this idea, \cite{salt:duci:2005} and \cite{bats:zogr:2011} explicitly model the strength of spatial dependence in the covariates to guide the selection of the spatial extent and weighting of neighboring covariate values to be used in the classification rule.   In cases where existing spatial classification methods do make use of the class of neighboring locations \citep{klei:pres:1992, pres:1996}, the dependence is not directly modeled.  Instead, the classes of neighboring locations are used to select the spatial extent of neighboring covariate values that are used as inputs to the classification rule. These final two methods, to our knowledge, are also the only existing Bayesian spatial classifiers.  These spatial approaches to classification have clear utility in remote sensing applications when entire scenes (images) are completely unobserved and spectra (covariates) are often strongly informative and exhibit strong spatial dependence.  However, in cases where only some pixel classes are missing, SGLM/SGLMM-based classifiers, which make direct use of both neighboring class information and covariates, is desirable.  

The primary goals of this paper are to formally define spatial binary classifiers based on the probit versions of the SGLM and SGLMM and compare these two classifiers in terms of the complexity of the underlying model and the robustness to misspecification of the underlying model.  As we discuss below, the probit SGLM and SGLMM have been used in the literature to model spatially-dependent binary data.  However, we are not aware of existing studies exploring model misspecification in the classification/prediction setting and thus this discussion is the primary contribution of this paper.  The formalism we introduce to define spatial binary classifiers based on the SGLM/SGLMM allows us to readily compare the performance of these methods to other spatial and non-spatial binary classifiers, a secondary goal of the paper.  We provide a comprehensive review of these alternative methods in an appendix (Appendix \ref{se:altclass}).

\section{SGLM and SGLMM Classifiers}\label{se:spaceclass}

\subsection{Spatial Generalized Linear (Mixed) Models}\label{se:SGLMs}

Following the seminal work of \cite{digg:tawn:moye:1998}, the generalized linear mixed model (GLMM) has become the go-to framework for modeling spatially-dependent phenomena for which Gaussian distributional assumptions are inappropriate.  Through the introduction of spatially-dependent random effects within a generalized linear model \citep{mccu:neld:1989},  standard models for both continuous and discretely-indexed Gaussian data (e.g., Gaussian processes, spatial autoregressive models) can be readily adapted to the non-Gaussian data situation.  (See \citealp{paci:2007}, for several examples, and \citealp{hugh:hara:2013}, for a more recent approach to dimension reduction and alleviation of confounding in these models.)   We review the SGLMM for dichotomous spatial data below in order to make connections between specific models that have appeared in the literature and to provide a framework for defining spatial classification methods.

Let $\bfY = (Y_1, \dots, Y_n)'$ be an observable binary response variables associated with locations $\bfs = (\bfs_1, \dots, \bfs_n)'$, which are a subset of either a continuous or discrete spatial domain.  
The standard generalized linear model (GLM) is specified through three components: the random component, the link function, and the systematic component.  The SGLMM can be written using these components as well.  For the random component of the model, we assume that conditional on unknown parameters $\bfbeta$ and $\bfnu$, the $Y_i$'s are independent and that
\begin{equation}
Y_i|\bfbeta, \bfnu \sim \mbox{Bin}(1, p_i), \label{eq:randcomp}
\end{equation}
where $p_i = P(Y_i = 1 | \bfbeta, \bfnu)$.  For the link function, we let
\begin{equation}
g(p_i) \equiv \eta_i, \label{eq:link}
\end{equation}
where generally we assume $g(\cdot)$ is a one-to-one, monotone, continuous, and differentiable function.  For binary response variables, common link functions are the logit and probit functions, $g(p_i) = \log\left({p_i}/({1-p_i})\right)$ and $g(p_i) = \Phi\inv(p_i)$, respectively, where $\Phi()$ is the standard normal cumulative distribution function and $\Phi^{-1}()$ denotes its inverse.  Finally, the systematic component of the model is
\begin{equation}
\eta_i = \bfx_i'\bfbeta + \nu_i,\label{eq:systcomp}
\end{equation}
where $\bfx_i$ is an $\ell \times 1$ vector of covariates measured at location $\bfs_i$, $\bfbeta$ is an $\ell \times 1$ vector of regression coefficients, and $\bfnu = (\nu_1, \dots, \nu_n)^{\prime}$ is a spatial random effects vector with components associated with each spatial location.  

By definition, the inclusion of random effects in the systematic component of the model (\ref{eq:systcomp}) makes the model a GLMM.  It also allows for the introduction of spatial dependence through the dependence structure of $\bfnu$.  In particular, we take
\begin{equation}
\bfnu | \bftheta  \sim \mathcal{N}(\bfzero, \bfSigma(\bftheta)), \label{eq:SRE}
\end{equation}
where $\bfzero$ is an $n \times 1$ vector of zeros and $\bfSigma(\bftheta)$ is the $n \times n$ spatial covariance matrix parameterized by $\bftheta$.  (For example, the spatial covariance matrix could correspond to a Mat\'ern covariance function or a spatial autoregressive model.)   As \cite{bane:carl:gelf:2004} point out, $\bfSigma(\bftheta)$ would not have a nugget, so that we can write
\begin{equation}
\bfSigma(\bftheta) = \theta_1K(\bftheta_2), \label{eq:spatdep}
\end{equation}
where $\bftheta = (\theta_1, \bftheta_2)'$ and $K(\bftheta_2)$ is an $n \times n$ spatial dependence matrix parameterized by $\bftheta_2$. Although not always the case, $\theta_1$ is often taken to be the common variance parameter among the $\nu_i$s and $K()$ is a spatial correlation function.  

The probit link function is often used in GLMs/GLMMs in the Bayesian setting since the well-known latent-variable representation of the probit GLM facilitates model fitting via the Gibbs sampler \citep{albe:chib:1993}.  This same latent variable representation also suggests an alternative GLM for dichotomous spatial data (compared to the spatial GLMM described above) that has been used in the literature, so we briefly review it here.  Introducing a collection of latent variables $\bfZ = (Z_1, \dots, Z_n)^{\prime}$ associated with each spatial location, we take
\[
Y_i = \begin{cases} 1, Z_i \ge 0,\\
0, Z_i < 0
\end{cases}
\]
where 
\begin{equation}
Z_i = \bfx_i'\bfbeta + \nu_i + \epsilon_i,\label{eq:ACSGLMM}
\end{equation}
$\epsilon_i \overset{iid}{\sim} \mathcal{N}(0, 1)$, and, as above, $\bfnu|\bftheta  \sim \mathcal{N}(\bfzero, \bfSigma(\bftheta))$.  We can see that this is simply an alternative representation of the probit SGLMM described previously since
\[
p_i = P(Y_i = 1|\bfbeta, \bfnu) = P(Z_i \ge 0|\bfbeta, \bfnu) = \Phi(\bfx_i'\bfbeta + \nu_i).
\]
Notice, however, that  
\begin{equation}
(\bfnu + \bfepsilon)|\bftheta \sim \mathcal{N}(\bfzero, {\bfSigma}^*(\bftheta)),\label{eq:spaterror}
\end{equation} where ${\bfSigma}^*(\bftheta) = \bfI + \bfSigma(\bftheta)$ and $\bfI$ is the $n$-dimensional identity matrix.  If ${\bfSigma}^*(\bftheta)$ is equal to $\bfSigma(\bftheta)$ instead (i.e., dropping the identity matrix), we have a spatial GLM (SGLM) as opposed to an SGLMM.  That is, there are no random effects in the systematic component of the GLM since the $\nu_i$s now take the place of the random component of the GLM in the latent variable representation of the model.  For this SGLM, we write 
\begin{equation}
Z_i = \bfx_i'\bfbeta + \nu_i.\label{eq:ACSGLM}
\end{equation}
With this specification, the $p_i$s are no longer found using the traditional inverse probit link function (the independent standard normal distribution). Instead, the joint probabilities can be found using the multivariate normal distribution.

Both the probit SGLMM and probit SGLM have been used in the literature.  Following \cite{deol:2000}, \cite{higg:hoet:2010} and \cite{berr:cald:2012} considered the SGLM, which is sometimes referred to as a clipped Gaussian process.  More recent work by \cite{schl:hoet:2013} makes use of the SGLMM for the ordered-category response variable case.  Given that both models are used in the literature, we briefly discuss the implications of their seemingly slight differences.  Comparing the probit SGLMM and SGLM described above, we note that while the $Y_i$s are conditionally independent under the SGLMM since the dependence is introduced through the dependence structure of the latent $\nu_i$s (\ref{eq:randcomp}), the $Y_i$s are not conditionally independent in the SGLM.  To see this, consider the \textit{conditional} likelihood function, $L(\bfbeta, \bfnu) = P(\bfY = \bfy \vert \bfbeta, \bfnu)$, for the SGLMM, which can be factored into $n$ components:
\begin{align*}
L(\bfbeta, \bfnu) &= \prod_{i=1}^n \left(\Phi(\bfx_i'\beta + \nu_i)\,I(y_i = 1) + (1 - \Phi(\bfx_i'\beta + \nu_i))\,I(y_i = 0) \right)   \\
& = \prod_{i=1}^n \int_{\mathcal{A}_i} \phi(z_i; \bfx_i'\bfbeta + \nu_i, 1)\, dz_i,
\end{align*}
where $\phi(\cdot; m, v^2)$ is the density function of the univariate normal distribution with mean $m$ and variance $v^2$ and 
\[
\mathcal{A}_i = \begin{cases} (-\infty, 0), &\mbox{ if }y_i = 0,\\
[0, \infty), &\mbox{ if }y_i = 1.
\end{cases}
\]  For the spatial GLM where there are no random effects in the systematic component of the model, the likelihood, $L(\bfbeta, \bftheta) = P(\bfY=\bfy \vert \bfbeta, \bftheta),$ is the integral over a multivariate normal distribution: 
\[
P(\bfY = \bfy|\bfbeta, \bftheta) = \int_{\mathcal{A}} \phi_n(\bfZ; \bfX\bfbeta, \bfSigma(\bftheta)) d\bfZ,
\]
where $\phi_n(\cdot, \bfm, \bfV)$ is the density function of the $n$-dimensional multivariate normal distribution with mean $\bfm$ and covariance $\bfV$, 
\[
\bfX = \left [\begin{matrix}
\bfx_1'\\
\vdots\\
\bfx_n'
\end{matrix}\right],
\]
and $\mathcal{A} = \mathcal{A}_1 \times \cdots \times \mathcal{A}_n$.  

Since the primary focus of this paper is spatial classification and model comparison, we defer discussion of model fitting to Appendix \ref{se:DA}.  However, we mention here recent approaches by \cite{hugh:hara:2013} and \cite{hank:etal:2015} to fitting a low-rank SGLMM for areal and geostatistical data, respectively, in the presence of spatial confounding.  It is well-known that when covariates are spatially dependent that coefficient estimates can be biased and variances inflated \citep[e.g.,][]{reic:etal:2006, hugh:hara:2013, hank:etal:2015}.  However, the impact of spatial confounding on prediction of the response variable has not been explored.  While examining this more fully is beyond the scope of this paper, we do compare the SGLM- and SGLMM-based classifiers' performance when the covariates are spatially dependent in Section \ref{se:simstudy}.  We discuss the \cite{hugh:hara:2013} parameterization of the SGLMM more fully in Appendix \ref{se:lowrank} and make use of their approach as an additional SGLMM classifier in Section \ref{se:comparisons}.

While the probit SGLMM and SGLM described above are nearly identical -- they differ only in terms of the inclusion of a identity matrix component in the model for ${\bfSigma}^*(\bftheta)$ -- it is not immediately clear how the models compare in terms of their predictive performance.  The difference between the models is reminiscent of a nugget effect, which is frequently used in geostatistical models and its inclusion is known to influence both point predictions and corresponding uncertainty statements.  Before conducting a comparison of the two models, we first describe how these models can be used to define spatial classifiers.  

\subsection{Model-Based Classification}\label{se:class}
In a modeling framework, imputing unknown response variables is often viewed as a prediction problem.  In the case where the unknown variable is categorical,  imputation is often viewed as a classification problem and many model-based and algorithmic classification methods exist.   To better relate the prediction methods from the SGLMM and SGLM to the other classification methods and to define a common classification rule among the various methods, we describe and build on the traditional classification problem to define decision functions for the SGLM and SGLMM.

\subsubsection{The Classification Problem}\label{se:classprob}

In the canonical classification problem, we have a collection of paired observations $\{(y_i, \bfx_i); i = 1, \dots, n\}$, where $y_i$ is the observed value of a binary response variable $Y_i$ and $\bfx_i$ is  an $\ell \times 1$ vector of covariates used as inputs in determining decision function boundaries for the classes.  The binary response variable is an indicator identifying the class to which the set of observed inputs belong.  Although many classification methods can be generalized to the multi-category/class setting, we restrict our description of these methods to the binary setting in which there are two classes, $C_0$ and $C_1$, where $C_j$ represents the class of observations where $Y = j$.  Of the observations $\{(Y_i, \bfx_i); i = 1, \dots, n\}$, $n_0$ fall into class $C_0$ and $n_1$ fall into class $C_1$, and $n_0 + n_1 = n$.  

A classification method defines a decision function $\delta(\bfomega)$, where $\bfomega$ is the set of applicable inputs, model parameters, and in the spatial setting, surrounding observations.  Based on this decision function, we can define a classification rule for an unobserved $Y$ that we denote generically by $Y^0$ with corresponding covariate information $\bfx^0$.  As described in \cite{hast:etal:2001}, the Bayes classifier for a 0-1 loss function is
\begin{equation}
y^0_{pred} = \arg\max_{j} P(C_j | \bfomega),\label{eq:bayesclass}
\end{equation}
where $P(C_j|\bfomega) = p_j(\bfomega)$ is the conditional probability of class $C_j$ given the inputs $\bfomega$. In other words, the optimal classification of $Y^0$ is the most likely class given $\bfomega$.  In the two-category case, we can rewrite (\ref{eq:bayesclass}) as
\begin{equation}
y^0_{pred} = \begin{cases}
1\mbox{, if }  \delta(\bfomega) > 1 \\
0\mbox{, otherwise}
\end{cases}. \label{eq:class}
\end{equation}
for $\delta(\bfomega) =   p_1(\bfomega)/p_0(\bfomega).$

For model-based classification, the conditional probabilities (and, hence decision functions) depend on unknown parameters denoted by $\bfxi$, which must be estimated from the observed data.  We write $\bfomega_{\bfxi}$ to explicitly capture the inclusion of the parameters in the collection of inputs.  In a frequentist setting, the unknown parameters, $\bfxi$, are often set equal to the maximum likelihood estimates (MLE), so that 
\[
\hat{p}_j(\bfomega) \equiv p_j(\bfomega_{\hat{\bfxi}}),
\]  where $\hat{\bfxi}$ is the MLE of $\bfxi$.  We refer to the classification rule given by (\ref{eq:class}) with $\delta(\bfomega)$ set equal to $\hat{\delta}(\bfomega) =  \hat{p}_1(\bfomega)/\hat{p}_0(\bfomega)$ as the \textit{maximum likelihood classifier}.   Alternatively, there are two Bayesian estimation approaches we consider.  First, we can set $\bfxi$ equal to its posterior mean, 
\[
\tilde{p}_j(\bfomega) \equiv p_j(\bfomega_{\tilde{\bfxi}}),
\]
where $\tilde{\bfxi} = E[\bfxi \vert \bfy]$.  In this case, the classification rule is given by  (\ref{eq:class}) with  $\delta(\bfomega)$ replaced by $\tilde{\delta}(\bfomega) = \tilde{p}_1(\bfomega)/\tilde{p}_0(\bfomega)$.  This classification rule can be viewed as the Bayesian analog of the maximum likelihood classifier, and we refer to it as the \textit{posterior mean classifier}.  An alternative Bayesian approach is to marginalize over the posterior distribution of $\bfxi$ and take
\begin{equation}\label{eq:postpredclass}
\bar{p}_j(\bfomega) \equiv \int_\Xi p_j(\bfomega_{\bfxi})\pi(\bfxi|\bfy)d\bfxi, 
\end{equation}
where $\pi(\bfxi \vert \bfy)$ is the posterior distribution of $\bfxi$ given $\bfy$ defined on the parameter space $\Xi$.  As before, we let $\bar{\delta}(\bfomega) =  \bar{p}_1(\bfomega)/\bar{p}_0(\bfomega)$ replace $\delta(\bfomega)$ in (\ref{eq:class}) and refer to the resulting classification rule as the \textit{posterior predictive classifier}.

\sloppypar{In practice, we evaluate the integral in (\ref{eq:postpredclass}) via Monte Carlo integration.  For example, for the Bayesian non-spatial GLM,  we approximate $\bar{p}_1(\bfomega)$ by drawing a realization of \,$Y^{0[t]} \sim \mbox{Bernoulli}(p_1(\bfx^0, \bfbeta^{[t]}))$ for $t=1, \dots, T$ and setting $\bar{p}_1(\bfomega) = \sum_{t=1}^T I(Y^{0[t]} =1)/T$, where $\bfbeta^{[t]}$ are the draws from the posterior distribution of $\bfbeta$,  and $T$ is the number of draws from the posterior distribution.  In the following subsections, our focus is on defining decision functions as it provides a unifying framework for both the model-based classifiers discussed below and the alternative classifiers discussed in Appendix \ref{se:altclass}.  }

\subsubsection{Spatial Classification}\label{se:spatialclass}
While the decision functions based on the non-spatial GLM's rely only on covariates and regression coefficients (see Appendix \ref{se:nonspatialclass} for details),  for the spatial case,  the decision function $\delta(\bfomega)$ also depends on the  categories of the surrounding observations. Because of this dependence,  it is necessary to define a joint distribution for the latent variables introduced in equation (\ref{eq:ACSGLMM}) for both the observations and the focal location.  Note that we generally define the same decision functions for the probit SGLMM and SGLM, since marginally, these models only differ in the definition of ${\bfSigma}^*(\bftheta)$ in (\ref{eq:spaterror}). 

Let $\bfZ^0 = (Z^0, \bfZ')'$ be the $(n + 1) \times 1$ vector of latent variables, where $Z^0$ is the latent variable for the unobserved location and $\bfZ$ is the $n \times 1$ vector of latent variables for the observations.  Then, combining equations (\ref{eq:ACSGLMM}) and (\ref{eq:spaterror}), 
\begin{equation}\label{eq:JointLatent}
\bfZ^0 \sim \mathcal{N}\left(\bfX^0\bfbeta, \bfSigma^{*0}(\bftheta)\right),
\end{equation}
where $\bfX^0 = (\bfx^0, \bfX')'$, 
\[
\bfSigma^{*0} = \left[ \begin{matrix} 
\sigma^0(\bftheta) &  \bfsigma(\bftheta)'\\
 \bfsigma(\bftheta) & \bfSigma^*(\bftheta)
 \end{matrix}\right],
\]
$\sigma^0(\bftheta)$ is the variance of $Z^0$, and $\bfsigma(\bftheta)$ is the $n \times 1$ vector representing the covariance of $Z^0$ and $\bfZ$.  
It follows that the distribution for the latent variable at an unobserved location is
\[
Z^0 |\bfx^0, \bfX, \bfbeta, \bftheta, \bfZ \sim \mathcal{N}(\mu_{Z^0}, \sigma_{Z^0})
\]
where 
\begin{align}
\mu_{Z^0} &= \bfx^{0\prime}\bfbeta + \bfsigma(\bftheta)'\left({\bfSigma}^*(\bftheta)\right)\inv (\bfZ - \bfX'\bfbeta),\label{eq:predmean}\\
\sigma^2_{Z^0} &= \sigma^0(\bftheta) - \bfsigma(\bftheta)' \left({\bfSigma}^*(\bftheta)\right)\inv \bfsigma(\bftheta).\label{eq:predvar}
\end{align}
We can easily include sampling from this distribution in the first step of the MCMC algorithm so that we can obtain draws from the posterior distribution of $Z^0$.  In this spatially-dependent case, $\bfomega = (\bfx^0, \bfX, \bfbeta, \bftheta, \bfZ)$ and
\begin{align*}
p_1(\bfomega) &\equiv p_1(\bfx^0, \bfX, \bfbeta, \bftheta, \bfZ) \\&= P(Z^0 > 0 |\bfx^0, \bfX, \bfbeta, \bftheta,  \bfZ) \\
&=  \Phi\left(\frac{\mu_{Z^0}}{\sqrt{\sigma^2_{Z^0}}}\right).
\end{align*}
Thus, the decision function at $\bfx^0$ is
\[
\delta_{SGLM}(\bfomega) \equiv \delta_{SGLM}(\bfx^0, \bfX, \bfbeta, \bftheta, \bfZ) = \frac{P(Z^0 >0 | \bfbeta, \bftheta, \bfZ)}{1 -P(Z^0 >0 | \bfbeta, \bftheta, \bfZ) } =  \frac{\Phi\left({\mu}_{Z^0}/\sqrt{{\sigma}^2_{Z^0}}\right)}{1 - \Phi\left({\mu}_{Z^0}/\sqrt{{\sigma}^2_{Z^0}}\right)}.
\]

\section{Comparison of the SGLM and SGLMM Classifiers}\label{se:compare}

\subsection{A Unified Model and Illustrative Comparisons}\label{se:unified_model}
To facilitate comparison of the SGLM and SGLMM classifiers, we consider the following representation of $\bfSigma^*(\bftheta)$ in (\ref{eq:spaterror}) of the SGLMM.  Recall that $\bfSigma^*(\bftheta) = \bfI + \bfSigma(\bftheta) = \bfI + \theta_1 K(\bftheta_2)$.   As discussed in Appendix \ref{se:DA}, we can augment the parameter space with an additional non-identifiable multiplicative factor on $\bfSigma^*(\bftheta)$.  This additional parameter facilitates MCMC mixing and allows us to consider a continuum of models indexed by a parameter $\kappa$.   Specifically, we  write
\begin{equation}\label{eq:withKappa}
\bfSigma^*(\bftheta) = \gamma^2( (1 - \kappa) \bfI + \kappa \bfSigma(\bftheta)),
\end{equation}
where $\gamma^2$ is a non-identifiable ``working" parameter and $\kappa \in [0,1]$ is the proportion of the marginal variance associated with the spatially-dependent variance component.  In other words, $\kappa \gamma^2 = \theta_1$.  The SGLM can now readily be viewed as a special case of the SGLMM.  Using this parameterization, as $\kappa \rightarrow 1$, the SGLMM becomes the SGLM and, as $\kappa \rightarrow 0$, the SGLMM becomes the independent probit model.  

\begin{figure}
\begin{center}
\includegraphics[width=\textwidth]{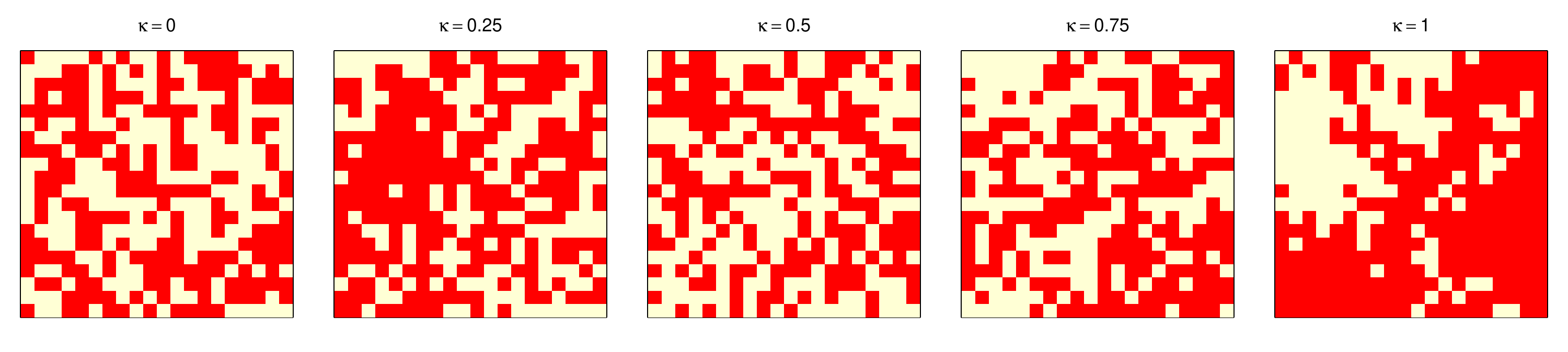}
\caption{Examples of binary data generated from the SGLM/SGLMM, where each data set is generated using a constant zero mean and varying levels of dependence.  The left is generated from the least spatially dependent model (i.e., independent probit) and the far right is generated from the most spatially dependent model (SGLM).}\label{fig:SimData}
\end{center}
\end{figure}

Although the SGLM and SGLMM are similar, data generated from these models can differ substantially in terms of the spatial structure.  For example, Figure \ref{fig:SimData} shows data sets generated under different SGLMMs, where for a fixed value of $\bftheta_2$ the data generated vary from completely independent to maximally spatially dependent (SGLM).    

For spatially-dependent binary variables, we expect neighboring observations to belong to the same class and wish to exploit this feature for improved classification.  To examine how the SGLM and SGLMM classifiers allow for this property, we provide two illustrations.  For both illustrations, we consider the SGLMM with $\bfSigma^*(\bftheta)$ given by (\ref{eq:withKappa}) and let $K(\bftheta_2)$ be defined by a conditionally autoregressive (CAR) spatial dependence structure \citep{bane:carl:gelf:2004} parameterized by $\rho$.  That is, $K(\bftheta_2) = (\bfI - \rho \bfW)^{-1}$, where $\bftheta_2 = \rho$ and $\bfW$ is a fixed, row-standardized $n \times n$ spatial neighborhood matrix.  In the first illustration where $n=2$, \[\bfW = \begin{bmatrix} 0 &1 \\ 1 &0 \end{bmatrix}.\] 

\begin{figure}
\begin{center}
\includegraphics[width=\textwidth]{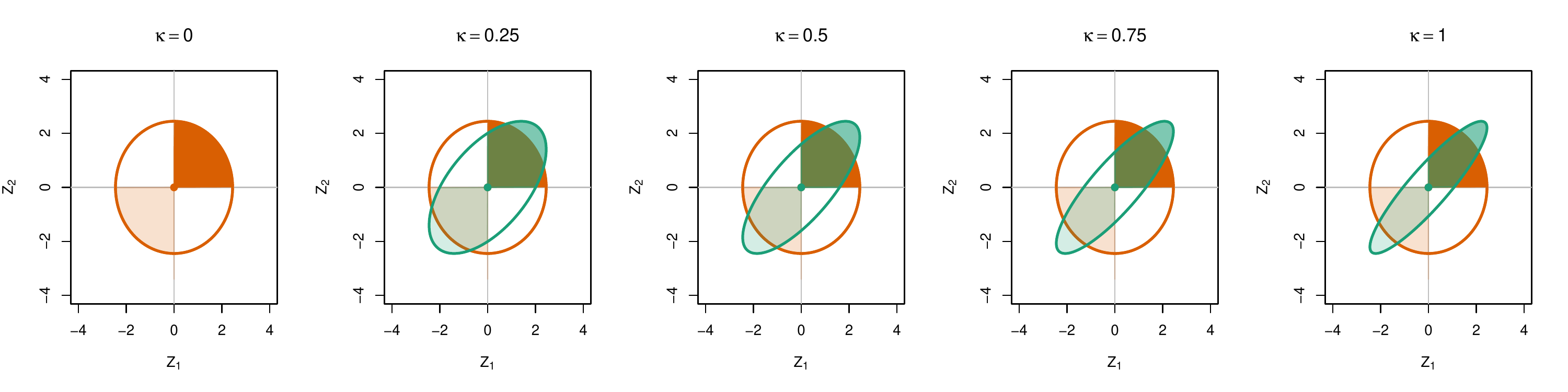}
\caption{Examples of 95 percent probability ellipses for two latent variables, $Z_1$ and $Z_2$, for the SGLM/SGLMM.  The shaded areas represent the probability of $Y_1$ and $Y_2$ being in the same class. }\label{fig:latentimages}
\end{center}
\end{figure}

First, consider the simple case where $n = 2$ so that we have observable binary variables $Y_1$ and $Y_2$ and focus on the event that both observations are from the same class (i.e., $\{Y_1 = 1, Y_2 = 1\}$ or $\{Y_1 = 0, Y_2 = 0\}$).  The probability of this event can be obtained from the joint distribution of the latent variables $Z_1$ and $Z_2$.   The plot on the left in Figure \ref{fig:latentimages} corresponds to the case when $Z_1$ and $Z_2$ are assumed to be conditionally independent given the values of other model parameters; this is the same as the standard independent probit regression model ($\kappa = 0$).  The orange circle represents the 95 percent ellipse of the joint probability distribution (i.e., the volume under the joint probability distribution is 0.95), which in this case is the circular bivariate normal distribution.  The center of this ellipse is marked by the orange point.  Note that the dark orange and light orange shaded areas within this ellipse represent $P(\{Y_1 = 1, Y_2 = 1\})$  and $P(\{Y_1 = 0, Y_2 = 0\})$, respectively, again assuming other model parameters are fixed.  This shaded ellipse is included in the remaining plots in Figure \ref{fig:latentimages} to aid in making comparisons.  The other images depict the same probabilities under different values of $\kappa$ for a fixed $\rho$.  These plots illustrate how as $\kappa$ increases, so do the probabilities of $P(\{Y_1 = 1, Y_2 = 1\})$  and $P(\{Y_1 = 0, Y_2 = 0\})$, even though the ``spatial dependence parameter," $\rho$, is fixed.  Therefore, for values of $0<\kappa < 1$, the strength of spatial dependence (i.e., probability of neighbors belonging to the same class) in the SGLMM does not simply depend on the value of $\rho$.

\begin{figure}
\begin{center}
\includegraphics[width=\textwidth]{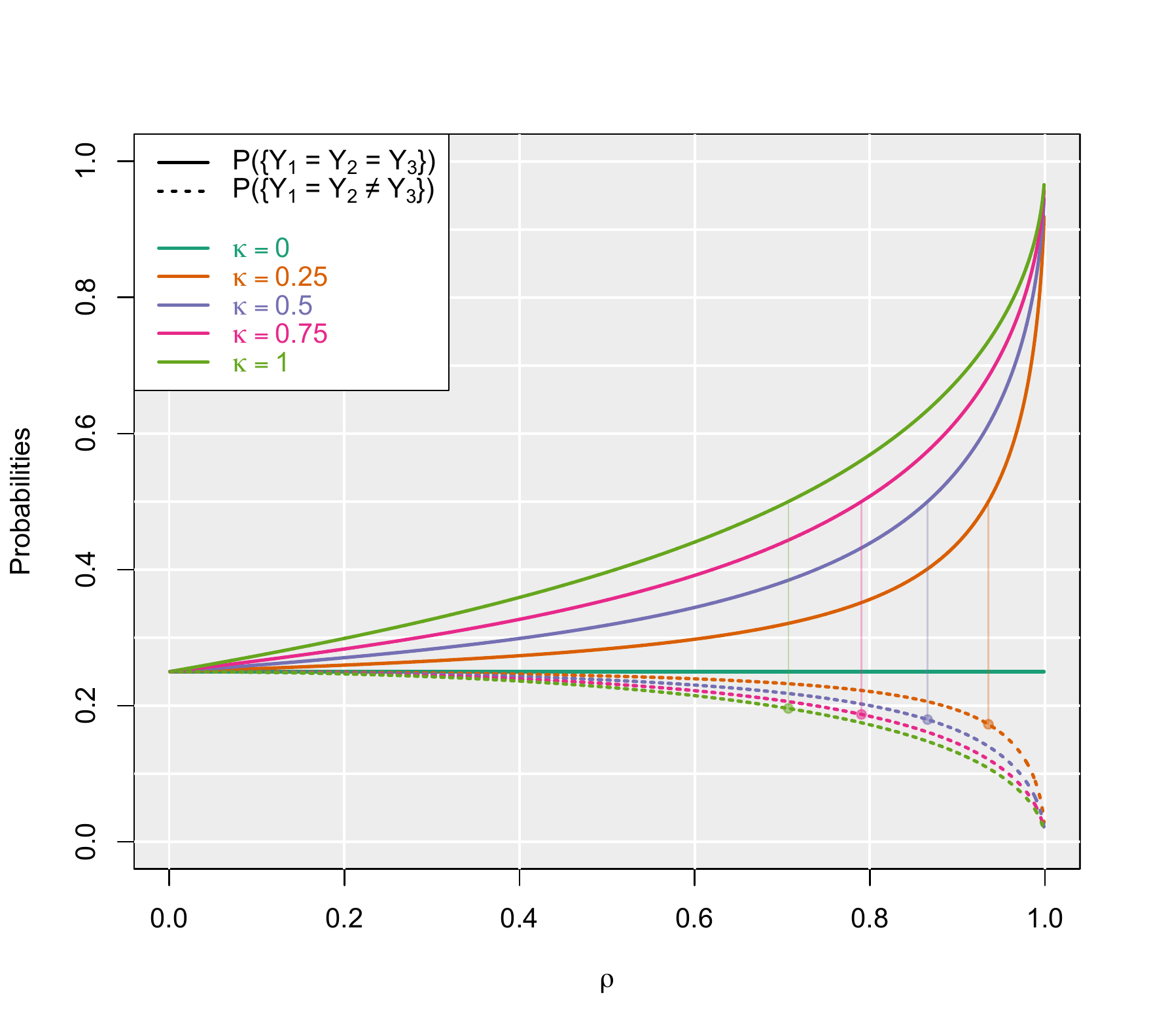}
\caption{Considering different combinations of classes for three locations, this plot shows how their probabilities change for different values of $\rho$ and $\kappa$.  The thick solid lines correspond to the probability that all three locations belong to the same class.  The dotted lines correspond to the probability that the third location belongs to a different class than the other two.  The thin solid lines connect these two probabilities for equivalent values of $(\rho, \kappa)$.  }\label{fig:probcurves}
\end{center}
\end{figure}

In the second illustration, $n = 3$ and
\[
\bfW = \begin{bmatrix} 0 & 1 & 0\\ 0.5 & 0 & 0.5 \\ 0 & 1 & 0 \end{bmatrix},
\]
so that the location of $Y_2$ is a neighbor to the locations of $Y_1$ and $Y_3$, but these two remaining locations are not neighbors of each other.    Again we are interested in the probability that these observations are the same: $P( \{ Y_1 = 0, Y_2 = 0, Y_3 = 0 \})$ and $P( \{ Y_1 = 1, Y_2 = 1, Y_3 = 1 \})$.  For shorthand, we represent this ``sameness probability" by $P( \{ Y_1 =  Y_2 = Y_3  \})$.  The solid lines in Figure \ref{fig:probcurves} show the value of this sameness probability for different combinations of $\rho$ and $\kappa$ when all other model parameters fixed.  As expected, for larger values of $\rho$ and $\kappa$, $P( \{ Y_1 =  Y_2 = Y_3  \})$ increases.  Notice, however, that there are several values of $(\rho, \kappa)$ that result in equal values of $P( \{ Y_1 =  Y_2 = Y_3  \})$.  In fact, Figure \ref{fig:probcurves} shows that  (0.935, 0.25), (0.866, 0.5), (0.790, 0.75), (0.707, 1) all result in $P( \{ Y_1 =  Y_2 = Y_3  \}) = 0.5$.  

Now consider the probability that $Y_3$ is different than $Y_1$ and $Y_2$: $P(\{ Y_1 = Y_2 \ne Y_3\}) = P( \{ Y_1 = 0, Y_2 = 0, Y_3 = 1 \}) + P( \{ Y_1 = 1, Y_2 = 1, Y_3 = 0 \})$.  The dotted lines in Figure \ref{fig:probcurves} represent these probabilities for different values of $\rho$ and $\kappa$, again with the other model parameters fixed.  Although the above combinations of $(\rho, \kappa)$ values resulted in equal values of $P( \{ Y_1 =  Y_2 = Y_3  \})$, they do not result in the equal values of $P(\{ Y_1 = Y_2 \ne Y_3\})$.  As shown by the points in Figure \ref{fig:probcurves}, the values of $P(\{ Y_1 = Y_2 \ne Y_3\})$ are close, but not equal.  This suggests that these models, and consequently classification rules derived from them, are robust to misspecification; that is, if the data were generated under the SGLMM/SGLM, the incorrect model can correct for the model misspecification through adjusting values of $\rho$ and/or $\kappa$.  To explore this in a more realistic setting where $n > 3$, in the following subsection we conduct a simulation study and compare the classifications for the SGLMM/SGLM under model misspecification.

\subsection{Robustness to Model Misspecification}\label{se:robust}

To more fully examine the differences between the SGLM- and SGLMM-based classifiers, we examine error rates for both classifiers under different generative models for the data.  Specifically, we are interested in determining if the model under which the data were generated performs the best in terms of classification, or if one model consistently outperforms the other.   Before describing the simulation study, we first define the method for computing the misclassification error rates we use for comparison.  

\subsubsection{Error Rates}

When developing and assessing classification methods, the observations are randomly divided into two subsets, the training and test data sets, of sizes $n_{train}$ and $n_{test}$, respectively.  This allows us to fit the model on which the classifier is based using the training data and evaluate the method's out-of-sample predictive ability.  When we compare classification methods, we consider training and test error rates, or the number of incorrect classifications among training and test data sets, divided by the sample size of each data set, respectively.  We expect training error to decrease as model complexity increases.  As our toy examples in the previous section illustrate, determining the relative complexity of the SGLM versus SGLMM is not straightforward -- we cannot simply count the number of parameters and conclude that the SGLMM is a more complex version of the SGLM.  On the other hand, test error rates allow us to explore potential overfitting.  


 In Section \ref{se:classprob} we described three model-based classifiers derived from different approaches for estimating the decision function: maximum likelihood, posterior mean, and posterior predictive.  Determining error rates  using maximum likelihood and posterior mean classifiers is straightforward.  However, for spatial classification,  determining training error rates based on the posterior predictive classifier is not.  Under the posterior predictive classifier, for the test data, we can simply use the latent $Z_j$, for $j = 1, \dots, n_{test}$, as sampled within the Gibbs sampler.  However, for the training data, the latent $Z_i$, for $i = 1, \dots, n_{train}$, are sampled within the Gibbs sampler given the observed $Y_i$.  In this case, to use the latent $Z_i$ as inputs, we must sample these values as if the $Y_i$ are unknown, otherwise we would have perfect prediction. 

For the Bayesian spatial probit regression model, we rely on the observed surrounding observations to provide information about the category of the unobserved locations.  Because of this, evaluating the training error is not straightforward.  We propose the following two approaches for the spatial probit posterior predictive classifier:
\renewcommand{\labelenumi}{\Alph{enumi}.}
\renewcommand{\labelenumii}{(\roman{enumii})}
\begin{enumerate}
\item One-at-a-Time Training Error:
\begin{enumerate}
\item Take samples $(\bfbeta^{[t]}, \bftheta^{[t]}, \bfZ_{\smneg i}^{[t]})$, for $t=1, \dots, T$, where $\bfZ_{\smneg i}^{[t]}$ is an $(n-1)\times 1$ vector of sampled $Z_j$ for $j= 1, \dots, i-1, i+1, \dots, n_{train}$ and sample a corresponding $Z^{[t]}_i \sim \mbox{N}(\mu_{Z_i}, \sigma^2_{Z_i})$ where 
\begin{align*}
\hat{\mu}_{Z_i} &= \bfx_i'\bfbeta^{[t]} + \bfSigma^*(\bftheta^{[t]})_{i,\smneg i}\left(\bfSigma^*(\bftheta^{[t]})_{\smneg i,\smneg i}\right)\inv ( \bfZ^{[t]}_{\smneg i} - \bfX_{\smneg i}\bfbeta^{[t]})\\
\hat{\sigma}^2_{Z_i} &= \bfSigma^*(\bftheta^{[t]})_{i,i} - \bfSigma^*(\bftheta^{[t]})_{i,\smneg i} \left(\bfSigma^*(\bftheta^{[t]})_{\smneg i,\smneg i}\right)\inv \bfSigma^*(\bftheta^{[t]})_{\smneg i, i} 
\end{align*}
and $\bfX_{\smneg i}$ is $\bfX$ with the $i^{th}$ row removed, and $\bfSigma^*(\bftheta^{[t]})_{j, \smneg k}$ is the $j^{th}$ row  of the estimated spatial correlation of $\bfZ$ with the $k^{th}$ column removed.  (Note that $\bfZ_{\smneg i}^{[t]}$ are the posterior samples obtained from the MCMC algorithm, however, the $Z_i^{[t]}$s are not the same as those sampled in the MCMC algorithm.)
\item Determine $\bar{p}_1(\bfomega) = \sum_{t=1}^T I(Z^{[t]}_i > 0)/T$ (suppressing the notation for the $i$-th observation in $\bar{p}_1(\bfomega)$), and let $Y^0_i$ be the predicted value of $Y_i$ using the posterior predictive classifier.
\item Repeat (i) and (ii) for all $i = 1, \dots, n_{train}$.
\item Compute the one-at-a-time training error: $\sum_{i = 1}^{n_{train}} I(Y^0_i \ne Y_i)/n_{train}$
\end{enumerate}
\item Joint Training Error:
\begin{enumerate}
\item Take samples $(\bfbeta^{[t]}, \bftheta^{[t]})$ for $t=1, \dots, T$ and sample a corresponding $\bfZ^{[t]} \sim \mbox{N}(\bfX\bfbeta^{[t]}, \bfSigma^*(\bftheta^{[t]}))$. (Note that the $\bfZ^{[t]}$ are not the same as those sampled in the MCMC algorithm.)
\item For $i = 1, \dots, n_{train}$, compute $\bar{p}(\bfomega) = \sum_{t=1}^T I(Z^{[t]}_i > 0)/T$ and let $Y^0_i$ be the predicted value of $Y_i$ using the posterior predictive classifier.
\item Compute the joint training error: $\sum_{i = 1}^{n_{train}} I(Y^0_i \ne Y_i)/n_{train}$
\end{enumerate}
\end{enumerate}
\renewcommand{\labelenumi}{\arabic{enumi}.}
\renewcommand{\labelenumii}{(\alph{enumii})}
Both the one-at-a-time and joint training errors allow for spatial dependence among the binary predictions/classifications through the latent random variable.   The joint training error allows for spatial dependence only through the spatial dependence structure of the latent variables, $\bfSigma^*(\bftheta)$.  In contrast, the one-at-a-time training error allows for spatial dependence  through $\bfSigma^*(\bftheta)$, but also allows for spatial dependence by conditioning on the current values of the latent random variables at nearby locations, $\bfZ_{\smneg i}^{[t]}$.

\subsubsection{Simulation Study}\label{se:simstudy}

For this simulation study, we generate a total of 45 data sets: three data sets for each of 15 unique scenarios.  Each data set has a grid size of 20$\times$20 and the scenarios vary in terms of covariate and spatial information.  To compare classification rates across different covariate values, we consider five versions of the linear component of the SGLM and SGLMM.  Table \ref{tab:simscenarios} outlines the different linear components under consideration and the corresponding parameter values.   `Intercept' is a model with only an intercept and is designed to examine classification rates when only spatial information is used to determine class.  `Simple-1' is a linear model with an intercept and a single covariate.  `Simple-2' is the same model as Simple-1, but with a larger coefficient value, designed to examine classification rates where there is strong covariate information.  `Multiple' is also designed for this purpose, but instead of large coefficients, has more covariates.  `Confounded' considers the case where the covariates are spatially dependent to compare the performance of the SGLM and SGLMM classifiers in the presence of spatial confounding.  For each of these five linear component models, we consider three different spatial dependence structures, thus giving 15 total scenarios.  

For all cases, $\bfSigma(\bftheta) \equiv \bfSigma(\rho)$, where $\bfSigma(\rho)$ is the covariance matrix defined by a conditionally autoregressive model introduced above with a second-order neighborhood structure such that locations that share a corner or an edge are taken to be neighbors.  For all data sets, we fix $\rho = 0.99$ and $\gamma^2 = 1$, and vary $\kappa$.  Therefore, we have three spatial dependence structures for each case: SGLMM-1 ($\kappa = 0.25$), SGLMM-2 ($\kappa = 0.5$), and SGLM ($\kappa = 1$).  Finally, covariate values are assigned by drawing the values from the distributions listed in Table \ref{tab:simscenarios}.

\rowcolors{2}{gray!18}{white}
\begin{table}
\begin{center}
{\small 
\begin{tabular}{|c|c|c|c|c|c|}
\hline
& \emph{Intercept} & \emph{Simple-1} & \emph{Simple-2} &  \emph{Multiple} & \emph{Confounded}\\
 & $\beta_0$ &$\beta_0 + \beta_1x_1$ & $\beta_0 + \beta_1x_1$ &  $\beta_0 + \beta_1x_1 + \beta_2x_2 + \beta_3x_3$ & $\beta_0 + \beta_1x_1$ \\\hline
$\beta_0$ &0.1 &0.1  & 0.1 & 0.1 & 0.1\\
$\beta_1$ & -- & $\neg\sqrt{2}$& $\neg\sqrt{8}$&  $\neg\sqrt{2}$ & $\neg\sqrt{2}$\\
$\beta_2, \beta_3$ & -- & -- & --& 2, 2 & -- \\
$x_1$ & -- & $\sim \mbox{U}(\neg0.5, 0.5)$& $\sim \mbox{U}(\neg0.5, 0.5)$& $\sim \mbox{U}(\neg0.5, 0.5)$ & $\sim \mathcal{N}(0, (I - 0.99W)^{\smneg1})$\\
$x_2, x_3$ & --& -- & -- & $\sim \mbox{U}(0, 0.5), \sim \mbox{U}(\neg0.5, 0)$ & --\\
$\rho$ &0.99& 0.99& 0.99 & 0.99 & 0.99 \\
$\kappa$ & 0.25, 0.5, 1 & 0.25, 0.5, 1 & 0.25, 0.5, 1& 0.25, 0.5, 1& 0.25, 0.5, 1\\\hline
\end{tabular}}
\caption{Parameter values and settings under which the various simulated data sets were generated.}\label{tab:simscenarios}
\end{center}
\end{table}

As is often the case with spatially-referenced data, missing observations occur in clumps of missing data, rather than scattered missing points.  Therefore, for this simulation study, we consider clustered test data, where the test data were taken in randomly selected clumps.  To do this, we randomly selected 25 of the locations and then randomly selected four of each location's eight neighbors (all locations, even those on the boundary, have 8 neighbors since we have access to the data over a larger region), removing any repeats and locations outside the region.  Using this approach, each data set has its own set of test locations, which make up approximately 27 percent of the locations.  We use the same prior distributions for all scenarios, namely, $\bfbeta \sim \mathcal{N}(\bfzero, 10 \times \bfI)$, $\rho \sim \mbox{U}(0, 1)$, $\kappa \sim \mbox{U}(0, 1)$, and for the working parameter in the data augmentation algorithms (see Appendix \ref{se:DA}) we use a scaled-inverse $\chi^2$ distribution, with 3 as the scale and degrees of freedom.  For each simulated data set, we fit the models using the algorithms described in Appendix \ref{se:DA}, run conservatively for 20,000 iterations to insure convergence.  We evaluated convergence using \cite{gewe:1992}'s convergence criteria.  

\begin{figure}
\begin{center}
\includegraphics[width=.6\textwidth]{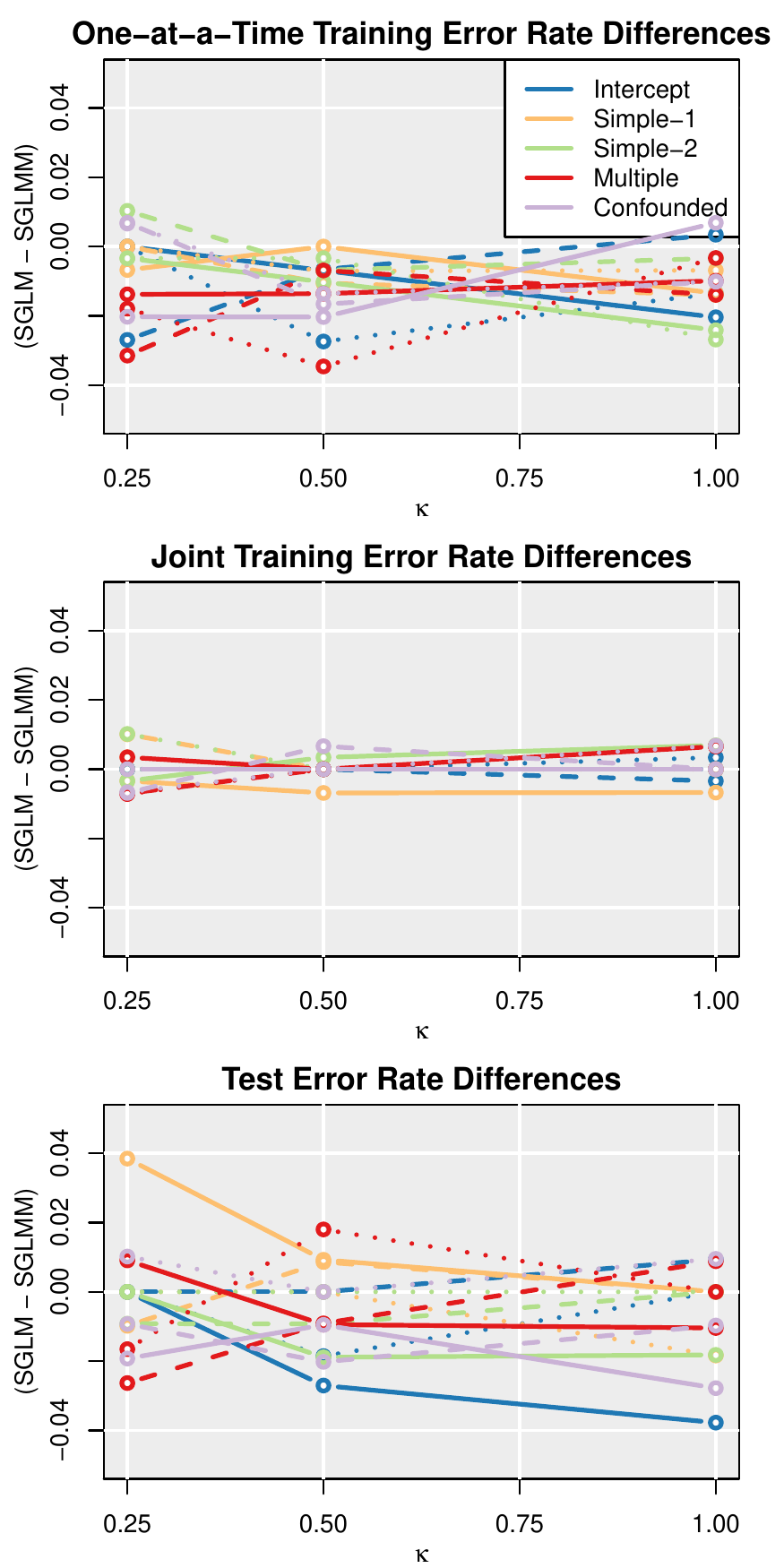}
\caption{Differences in the error rates between the SGLM and SGLMM classifiers.  Below 0 indicates the SGLM-based classifier performed better, while above 0 indicates the SGLMM-based classifier performed better.}\label{fig:diffs}
\end{center}
\end{figure}

We include exact error rates in Appendix \ref{se:simrates}, but refer here to Figures \ref{fig:allrates} and \ref{fig:diffs} to more clearly compare the two classifiers.  Figure \ref{fig:allrates} shows the misclassification error rates for all data sets, where the x-axis represents the amount of spatial dependence under which the data set was generated ($\kappa$).    As expected, this figure shows that in general, the classification rules tend to perform better when there is stronger spatial dependence and stronger covariate information; the scenarios producing the smallest error rates corresponds to the data sets generated with the large coefficient value (Simple-2) and the spatially-dependent covariates (Confounded), with maximal spatial dependence ($\kappa = 1$).  Furthermore, there is not much difference between the error rates for the SGLM- and SGLMM-based classifiers, regardless of which model was used to generate the data.  This is more clearly seen in Figure \ref{fig:diffs}, where the differences between the error rates of the SGLM and SGLMM are shown for each data set.  In this case, there is not a strong pattern across the scenarios indicating when a particular classifier might be preferred over another.  We take this as evidence of the robustness of the SGLM/SGLMM-based classifiers to model misspecification and find no evidence that the SGLMM is more complex than the SGLM despite its additional parameter.  For one-at-a-time training error rates, in this small sample of data sets, it does appear that the SGLM fits better than the SGLMM when the true model contains stronger spatial dependence.

\begin{figure}
\begin{center}
\includegraphics[width=\textwidth]{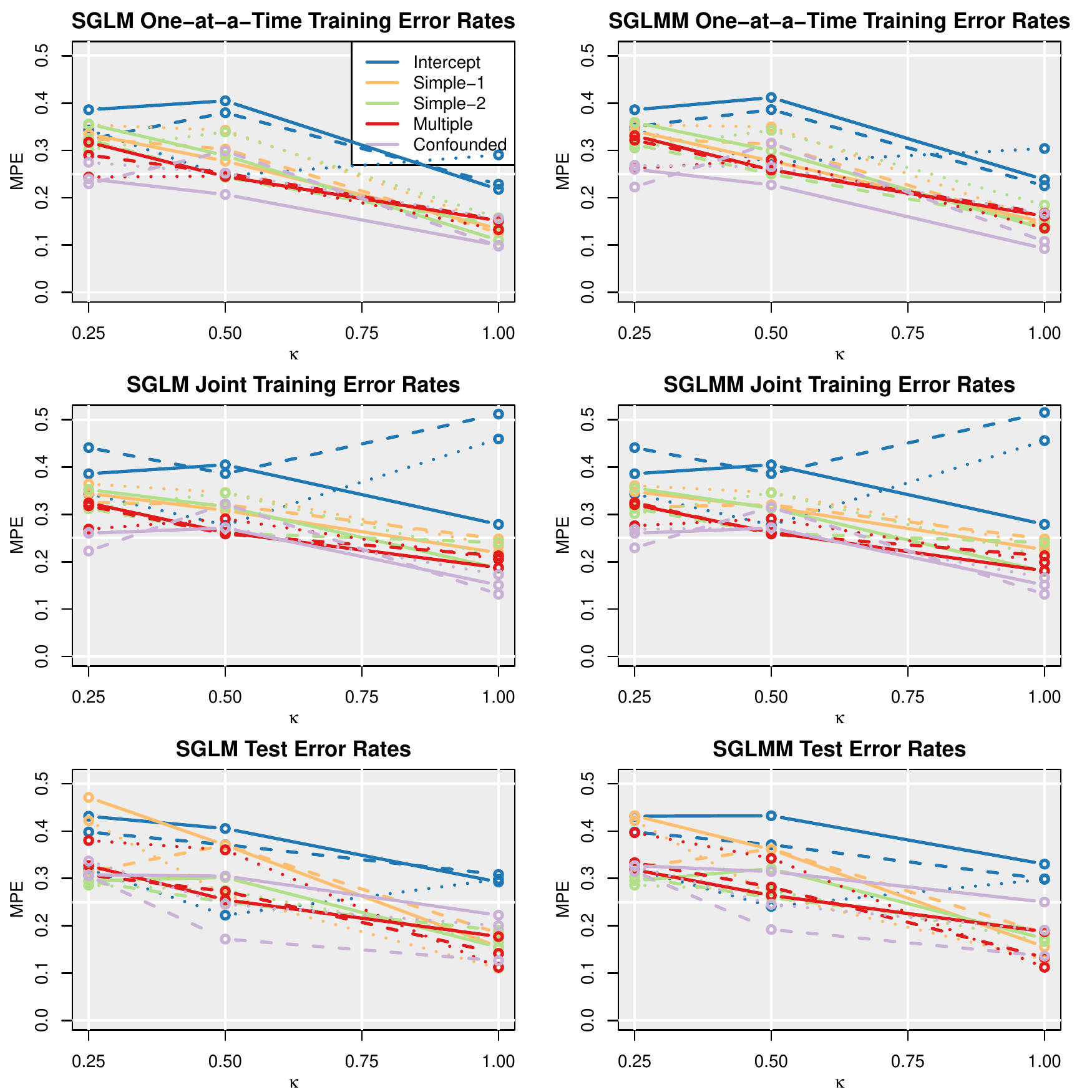}
\caption{Classification error rates for each data set.  The colors indicate the model from which the data are generated:  `Intercept' (green); `Simple-1' (blue); `Simple-2' (orange); `Multiple' (red); and `Confounded' (purple).  The three different line types represent the three different data sets for each scenario.  As a function of $\kappa$ (which together with the colors completes the specification of the model that generated the data), the top row summarizes the one-at-a-time training error rates, the middle row summarizes the joint training error rates, and the bottom summarizes the test error rates.  The left column contains the error rates when the SGLM is fit to the data, and the right column contains the error rates when the SGLMM is fit to the data.  }\label{fig:allrates}
\end{center}
\end{figure}

\section{Empirical Comparison of Classification Methods}\label{se:comparisons}

\subsection{Land Cover Application}

To illustrate the SGLM/SGLMM classifiers and other commonly used classification methods, we provide an empirical example using satellite-derived observations of land cover over Southeast Asia.  Over the last century, Southeast Asia has experienced much deforestation, where as much as 12 percent of original forests have been lost to other land uses \citep{munr:etal:2008}.  Researchers are interested in tracking the changing land covers to determine economic, geographic, social, and demographic factors that may contribute to deforestation.  Satellite measurements provide a record of historic land use, but in many instances, cloud cover and other weather events may prevent the satellite from obtaining a consistent measurement for determining land cover at every location.  

We used data from the National Aeronautics and Space Exploration (NASA)'s Moderate Resolution Imaging Spectroradiometer (MODIS) Land Cover Type Yearly Level 3 Global 500m (MOD12Q1 and MCD12Q1) data product for the year 2005.  Our region of interest covers the region bounded by 17$^\circ$ to 19$^\circ$N and 98$^\circ$ to 100$^\circ$E, which covers a portion of northwestern Thailand and a small part of Myanmar.  Because observations from this data are made at such a fine resolution, for computational convenience we collapsed the data into a 24 $\times$ 24 grid (each cell is approximately 8.8 km $\times$ 9.25 km) by selecting the most common land cover -- forest or non-forest -- in each grid cell.

We considered four covariates: elevation, distance to the nearest major road, distance to the coast, and distance to the nearest big city.  Elevation is measured in meters and distances are Euclidean and measured in degrees.  The covariates are standardized, meaning that there were no costs taken into account in calculating distance (e.g., distance calculations do not take into account the fact that it might take longer to go over mountains than go around them).  To find the covariate values for the grid cells, for each covariate, we used the median value of all the observed locations within the grid cell.

\begin{figure}
\begin{center}
\includegraphics[width=6in]{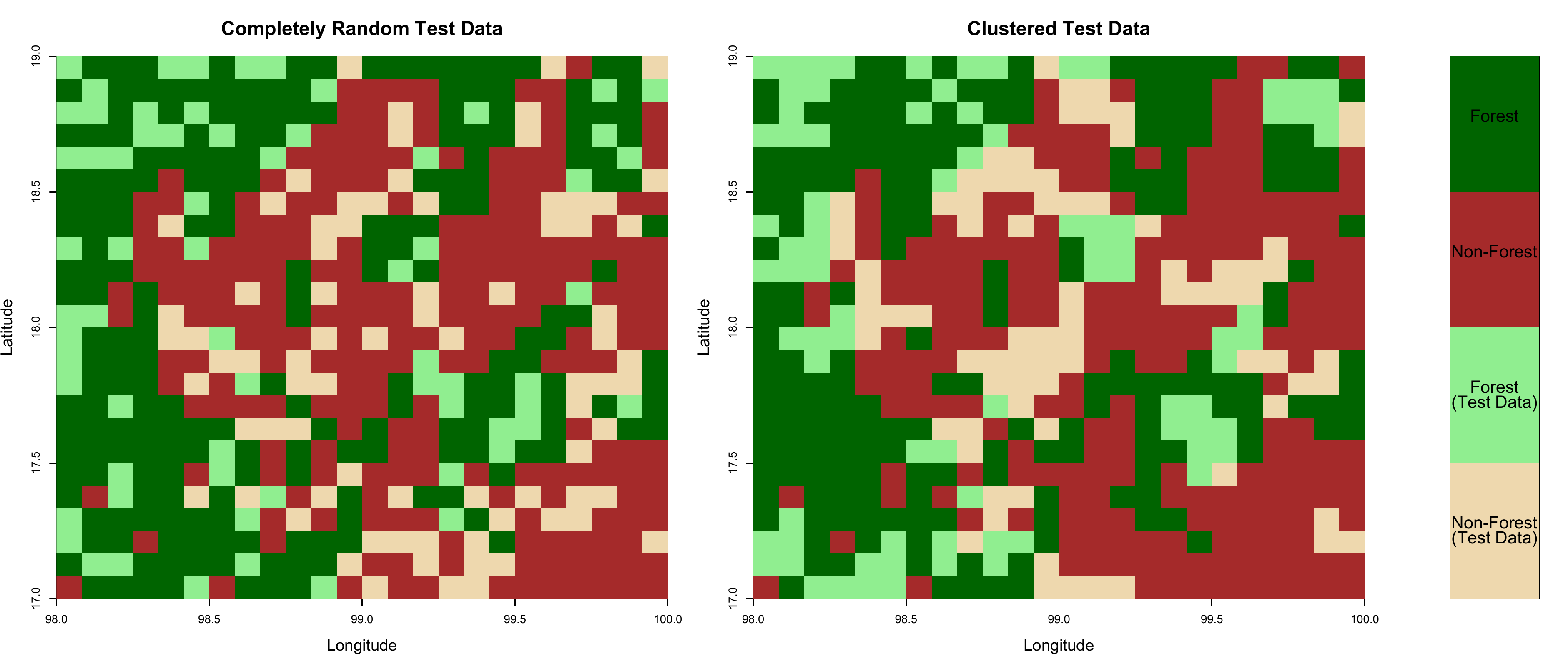}
\caption{Non-clustered (left) and clustered (right)  training (dark colors) and test (light colors) data sets. Forested locations are in green and non-forest locations are brown.}\label{fig:data}
\end{center}
\end{figure}

With satellite data, missing observations often occur in clumps of missing data, rather than scattered points of missing data.  Because of this, we randomly selected two test data sets: non-clustered and clustered.  For the non-clustered test data, we randomly selected $n_{test} = 144$ locations as test data and used the remaining $n_{train} = 432$ as training data.  For the clustered test data, we followed a similar method as described in Section \ref{se:simstudy}.  We randomly selected 36 locations and then randomly selected four of each location's eight neighbors (all locations, even those on the boundary, have 8 neighbors since we have access to the data over a larger region), removing any repeats and locations outside the region.  This resulted in $n_{test} = 159$ locations assigned to the clustered test data, and $n_{train} = 417$ assigned to the training data.  Figure \ref{fig:data} shows images of both the non-clustered and clustered  training and test data sets.  We examined training and test error rates for each of these test data sets; however, we also examined test error rates for two more test data sets for each type of missing data.

\subsection{Prior and Tuning Parameter Values}

The Bayesian probit models require prior distributions on parameter values.  For the independent probit model (i.e., $\bfSigma^*(\bftheta)$ is the $n \times n$ identity matrix in (\ref{eq:spaterror})), SGLM, and SGLMM, we assign prior distribution $\bfbeta \sim \mathcal{N}(\bfzero, 10\times\bfI)$, where $\bfbeta$ are the coefficients for the centered and scaled covariates.  For the working parameter in the data augmentation algorithms (see Appendix \ref{se:DA}), we use a scaled-inverse $\chi^2$ distribution, with $a_\gamma = 3$ degrees of freedom and $b_\gamma = 3$ as the scale parameter.  For the spatial dependence parameter in the SGLM and SGLMM, $\theta \sim \mbox{U}(a_\theta, b_\theta)$.  We used a conditionally autoregressive (CAR) dependence structure with a second-order neighborhood structure and set $a_\theta = 0$ and $b_\theta = 1$ (since the spatial neighborhood matrix, $\bfW$, is row-standardized, the maximal value of $\rho$ is 1 and we use a lower prior bound of 0 since we expect spatial dependence to be non-negative).  For the SGLMM, we assign $\kappa$ a $\mbox{Unif}(0,1)$ prior distribution.  To fit each of these models, we ran the MCMC for 120,000 iterations, at which point all chains are well-converged as determined by examining trace plots of the parameters and \cite{gewe:1992}'s diagnostic, and use the last 100,000 for prediction and determining error rates.

We follow the approach of \cite{hugh:hara:2013} for fitting the low-rank probit SGLMM (label ``Hughes and Haran" below).  Namely,  we use the eigenvectors of the Moran operator corresponding to the largest 10 percent of the eigenvalues for the values of $\bfPsi$ (see Appendix \ref{se:lowrank}).  We use a $\mathcal{N}(\bfzero, 100\bfI)$ as the prior distribution for the coefficients of $\bfPsi$.  

Support vector machines (SVM) and k-nearest neighbors (kNN) classification methods require tuning parameters, $\lambda$, $u$, and $k$ (see Appendix \ref{se:altclass}).  We obtain optimal values of these parameters using five-fold cross-validation on each training data and assigned the value of the associated parameter to be the value with the lowest cross-validation error (CVE).  Table \ref{tab:CVE} shows the chosen tuning parameter and associated CVEs  for each classification method and data set.

\rowcolors{2}{white}{white}
\begin{table} 
\begin{tabular}{c|c|cc|cc}
&  Tuning& \multicolumn{2}{c|}{\it Non-Clustered} & \multicolumn{2}{c}{\it Clustered} \\
Classification Method & Parameter &  Optimal Value & CVE & Optimal Value& CVE \\\hline
Linear SVM & $\lambda$ &0.39 & 0.2767&0.51& 0.3084  \\
Cubic SVM & $\lambda$ &0.98 & 0.2837 &0.8 & 0.2964  \\
Radial SVM & $\lambda$, $u$ & 0.515; 5.152& 0.2140  &0.875, 7.778 & 0.2410 \\
kNN-C & $k$ &5 & 0.2326 &12 & 0.2506 \\
kNN-G & $k$ & 5 & 0.1953& 3& 0.1880  \\
\end{tabular}
\caption{Optimal value of tuning parameters and the associated five-fold cross-validation errors for each classification method and data set used for the results in Table \ref{tab:classerr}.}\label{tab:CVE}
\end{table}

\subsection{Results}

We compared error rates for the model-based classifiers discussed in Section \ref{se:class} with additional classification methods described in Section \ref{se:altclass}:  linear, diagonal, and quadratic discriminant analysis classifiers (LDA, DLDA, and QDA, respectively); linear, cubic, and radial support vector machines (SVM); k-nearest neighbor (kNN) classifiers where neighbors are defined either in covariate space (kNN-C; the traditional approach) or geographic space (kNN-G); and four spatial classifiers from \cite{swit:1980} (``Switzer"), \cite{mard:1984} (``Mardia"), \cite{salt:duci:2005} (Spatial LDA), and \cite{pres:1996} (``Press").  We examine in-depth the training and test errors for a single test data set in Table \ref{tab:classerr}.  We also examine test error rates for more test data sets in Table \ref{tab:moretest}.  We considered two scenarios of missing data: non-clustered and clustered training/test data sets (see Figure \ref{fig:data}). 

\begin{table}[h!!]
\begin{center}
\begin{tabular}{llcccc}
& \multicolumn{1}{c}{\, }& \multicolumn{2}{c}{\bf  \emph{Non-clustered Data}} & \multicolumn{2}{c}{\bf \emph{Clustered Data}}\\
\multicolumn{2}{c}{\, \bf Classification Method} & { \bf Training }& \multicolumn{1}{c}{\bf Test}& {\bf Training }&  \multicolumn{1}{c}{\bf Test}\\ \hline
\\[-.1in]
\multicolumn{2}{l}{\bf Model-based Classifiers}&&&&\\[.05in]
&{\it SGLM} &&&&\\ 
& \, Posterior Mean &  0.1551 & 0.1667 & 0.1775 & 0.1447 \\ 
&\,  Posterior Predictive & 0.1551; 0.2731 & 0.1667 & 0.1775; 0.2998 & 0.1447\\[.05in]
&{\it SGLMM} &&&&\\ 
 & \, Posterior Mean & 0.1644 & 0.1736 & 0.1823 & 0.1447  \\ 
&\,  Posterior Predictive & 0.1829; 0.2894& 0.1667 & 0.1823; 0.3046 & 0.1447\\
&\, Hughes and Haran &  0.1296 & 0.2222 & 0.1367 & 0.1696   \\[.05in]
&{\it Probit - Bayesian}&&&&\\ 
&\,  Posterior Mean & 0.2824 & 0.2986&  0.2854 & 0.2767   \\ 
&\,  Posterior Predictive & 0.2824 &0.2986  & 0.2854  &  0.2767 \\[.05in]
&{\it GLM - Maximum Likelihood} &&&&\\ 
&\, Logistic &  0.2824 & 0.3056 & 0.2854 & 0.2767   \\ 
&\, Probit &    0.2824 & 0.2986 & 0.2854 & 0.2767  \\[.1in]
\multicolumn{2}{l}{\bf Additional Classifiers}\\[.05in]
&{\it Discriminant Analysis}&&&&\\ 
&\, LDA &  0.2778 & 0.2917 &  0.2878 & 0.2704  \\ 
&\, DLDA & 0.3611 & 0.3889 & 0.3549 &0.3522   \\ 
&\, QDA &  0.2593 & 0.2986 & 0.2638 & 0.2767  \\[.05in]
&{\it SVM}&&&&\\  
&\,  Linear SVM & 0.2847 & 0.3264  &0.2926 & 0.2956   \\ 
&\, Cubic SVM & 0.2685 & 0.2847   &0.2806 & 0.3396 \\
&\, Radial SVM &   0.0949 & 0.2639 &  0.0530 & 0.1950 \\[.05in]
&{\it k-Nearest Neighbors}&&&&\\ 
&\, kNN-C & 0.1597 & 0.2639 & 0.2062 & 0.2579 \\
&\, kNN-G &0.1204 & 0.1458 & 0.1055 & 0.2075\\[.05in]
& {\it Spatial Classifiers}&&&&\\
& \, Switzer &0.2616&0.2639&0.2566&0.2579\\
& \, Mardia &0.4074&0.4375&0.5348&0.5094\\
& \, Spatial LDA &0.3634 &0.3750&0.3741&0.4717\\
& \, Press &0.3912&0.3611&0.3933&0.4214\\[.1in]
\hline

 \end{tabular}
  \caption{Training and test errors for the SE Asia land cover data obtained using various classification methods. The posterior predictive errors for the training data list two errors: one-at-a-time (left) and joint (right).}\label{tab:classerr}
 \end{center}
 \end{table}

 \begin{table}[h!!]
\begin{center}
\begin{tabular}{llcccc}
& & \multicolumn{2}{c}{\emph{\textbf{Non-clustered Data}}} & \multicolumn{2}{c}{\emph{\textbf{Clustered Data}}} \\ 
\multicolumn{2}{c}{\textbf{Classification Method}} & \textbf{Test Set 1} &  \textbf{Test Set 2} & \textbf{Test Set 3} & \textbf{Test Set 4}\\ \hline\\[-.1in]
\multicolumn{2}{l}{\bf Model-based Classifiers}&&&&\\[.05in]
&{\it SGLM} &&&&\\ 
&\, Posterior Mean & \textbf{0.1806}  & \textbf{0.1806} & 0.1688 & 0.2318\\
 &\,  Posterior Predictive &  \textbf{0.1806}  &  \textbf{0.1806}  &  0.1688  &  0.2318     \\[.05in]
\hspace*{.01in}&{\it SGLMM}& \\
&\, Posterior Mean & \textbf{0.1806} & 0.1944 & 0.1688 & \textbf{0.2252}\\
 &\,  Posterior Predictive&  \textbf{0.1806}  &  0.1944  &  \textbf{0.1623}  &  \textbf{0.2252}     \\
 &\, Hughes and Haran &0.1875 & 0.1875& 0.1688 &0.2715 \\[.05in]
  &{\it Probit - Bayesian} &&&&\\ 
  &\, Posterior Mean & 0.2986 & 0.3056 & 0.2143 & 0.3245\\
  &\, Posterior Predictive &0.2917 & 0.3056 & 0.2013 & 0.3245\\[.05in]
 &{\it GLM - Maximum Likelihood} &&&&\\ 
& \, Logistic &   0.2986  &  0.3125   &0.2078  &  0.3311   \\ 
 & \, Probit &  0.2986  &  0.3056  &  0.2143  &  0.3245    \\[.05in] 
\multicolumn{2}{l}{ \textbf{Additional Classifiers}} & & &  \\ 
&{\it Discriminant Analysis}&&&&\\ 
& \, LDA&  0.2917  &  0.3264 &  0.2143  &  0.3179      \\ 
 &\, DLDA &  0.3681  &  0.4167&  0.3377  &  0.5033     \\ 
 &\, QDA&  0.2847  &  0.3194 &  0.2143  &  0.3510      \\[.05in] 
 &{\it SVM}&&&&\\  
& \, Linear SVM&  0.2986  &  0.3264  &  0.2143  &  0.3245     \\ 
 &\, Cubic SVM &  0.2917  &  0.3333   &  0.2727  &  0.3775   \\ 
 &\, Radial SVM &  0.2847  &  0.3056  &  0.2143  &  0.3510    \\[.05in]
 &{\it k-Nearest Neighbors}&&&&\\  
& \, kNN-C &  0.2708  &  0.2917 & 0.2792  &  0.3444     \\ 
 &\, kNN-G &  0.2361  &  0.2847   &  0.3117  &  0.2715   \\[.05in] 
 & {\it Spatial Classifiers}&&&&\\
& \, Switzer &  0.2639  &  0.3056 &  0.1753  &  0.2715     \\ 
 &\, Mardia &  0.4514  &  0.4583 &  0.4416  &  0.4437      \\ 
 &\, Spatial LDA &  0.4514  &  0.4583    &  0.4416  &  0.4437  \\ 
 &\, Press & 0.5556 & 0.5903 &  0.5584 &0.4636  \\[.05in]\hline
\end{tabular}
\caption{Test errors under various classification methods for multiple test data sets for the SE Asia land cover data. The bolded test errors are the smallest test errors across all classification methods for that particular test set.}\label{tab:moretest}
\end{center}
\end{table}

 We first compare training and test error rates for the first sets of test data, shown in Table \ref{tab:classerr}.  The training and test error rates tend to be similar for each classification method.  Furthermore, the non-clustered and clustered data sets have similar error rates.  Most of the error rates from non-spatial methods fall around 30 percent; however, the classification methods which make use of neighboring classes (i.e., the SGLM, SGLMM, and kNN-G classifiers) have much lower error rates.  

As in Section \ref{se:simstudy}, the SGLM and SGLMM error rates are similar, with the SGLM doing slightly better in this case.  
While not included here, we note that the reported error rates were similar when we used a geostatistical-type exponential covariance structure rather than a CAR model.  Therefore, we do not believe the differences in error rates are a consequence of the particular form of the spatial dependence structure and for computational convenience we prefer the CAR model.  

As expected, the SGLM and SGLMM joint training error rates are not as small as the one-at-a-time training error rates which has the added benefit of using of the observed classes of neighboring locations and not just allowing for spatial dependence in the residual of the latent variable.   The one-at-a-time training error rates are also more similar to the test error rates, implying that the one-at-a-time training error rates would be more useful in a practical classification setting.  

For the clustered test data, the SGLM and SGLMM-based classifiers have the lowest error rates.  kNN-G also uses neighboring classes and has small error rates relative to the other competing methods.  Although this classifier has smaller rates for the training data, it does not have the smallest error rates across the board.  Furthermore,  the error rates for this classifier are not fixed because of the random classification when there are ties.  For example, repeating this classification method several times for the non-clustered data, the test error rates for kNN-G ranged from 12 percent up to 21 percent.  Although including only the classes of neighboring locations improves the error rates over  those methods which do not make use of neighboring classes, also including covariates in the classifier -- as in the SGLM or SGLMM -- provides as low and more consistent error rates.

SVM classification methods offer more flexibility in the relationship between the inputs and the probability of the classes than a GLM.  The radial SVM classifier does quite well, and in fact has the smallest training error rates.  However, the test error rates are not as small as those for the spatial methods, suggesting that allowing for this more complicated structure among the inputs is not sufficient to produce better predictions when classifying out-of-sample data in space.  

The other spatial classifiers (``Switzer", ``Mardia", Spatial LDA, and ``Press") make use of spatial structure in the covariates/inputs.  Although the ``Switzer" classifier is competitive with LDA, QDA, and SVMs, the other spatial classifiers -- including the Bayesian classifier (``Press") -- perform rather poorly in comparison to the other classification methods.   This is perhaps due to the fact that the spatial structure of the covariates differs from that of the process of interest (forest/not forest).  In this case, we want to take advantage of the covariate information at the focal location, but not at the neighboring locations. 
Comparing the error rates of all these classification methods suggests that including inputs associated with the focal location as well as neighboring observations, as in the SGLM and SGLMM classifiers, leads to better classification methods.

Table \ref{tab:moretest} shows similar test error rates to those found in Table \ref{tab:classerr}. However, the SGLM- and SGLMM-based classifiers perform the best across the board.  We also note that the \citeauthor{hugh:hara:2013} approach which accounts for spatial confounding between the covariates and the spatial random effect fits almost as well as the full SGLM- and SGLMM-based classifiers, but not quite.  We assume this is because of the dimension reduction which smoothes out some of the spatial dependence,  which may be important in making use of neighboring pixel classes.  

In terms of computation time, the SGLM has a large computational advantage over the SGLMM, as it has one less parameter and, for the CAR dependence structure, it does not require inverting the $n \times n$ covariance matrix.  For the 120,000 iterations run for this application, the SGLM took 17.74 CPU hours running on a single core of an Intel Xeon CPU E7-8837 \@ 2.66GHz, while the SGLMM took 50.50 CPU hours using up to 18 cores where possible.  Fitting these classifiers are much more computationally intensive than the others described in this paper, where the non-Bayesian methods are nearly instantaneous.  While perhaps some would consider this large difference in computational time, we note that it may be worth it.  In our example from Table \ref{tab:classerr}, the SGLM classified 6 percent more of the locations correctly than the next best classifier (kNN-G).  Furthermore, since the SGLM- and SGLMM-based classifiers are based on an underlying statistical model, we can obtain measures of uncertainty that the algorithmic-based classification methods cannot.  Lastly, we note that we ran the MCMC algorithms longer than is likely needed in order to be conservative.  If fitting time was a major concern, a less conservative strategy might be warranted, such as the dimension reduced ``\citeauthor{hugh:hara:2013}" SGLMM, which required only 1.4 CPU hours.

\section{Discussion}

In this paper, we discuss Bayesian classifiers for spatial data based on the Bayesian probit SGLM and SGLMM.  Although these models are similar, we attempt to provide some insight into their differences.  Through two illustrative examples and a simulation study, we find that classification rules based on these models are fairly robust to model misspecification.  Another contribution of this paper is a comprehensive review and empirical comparison of alternative classifiers in the literature.  While the SGLM/SGLMM-based classifiers outperform others in the literature for our land cover application, we caution against drawing general conclusions from these findings.  For other applications, the performance of the classifiers may differ.  We hope that our overview of spatial classification methods will be of use to researchers in selecting candidate methods for other applications.

We focused on building spatial classifiers using a probit SGLM and SGLMM, but other spatial statistical models for discrete data may also suggest classification rules that include neighboring location classes and covariates as inputs.  For example, \cite{hoet:etal:2000} developed a spatial classifier based on the autologistic model.  A comparison of alternative spatial model-based classifiers is left to future work.  In addition, we note that many of the other classification methods cited in this work allow for classification of more than two categories.  \cite{albe:chib:1993} extend their latent variable probit model to ordinal and categorical response variables.  Using this framework, the SGLM and SGLMM have been extended to model these types of response variables in a spatial setting \citep[see, for example,][]{berr:2010, schl:hoet:2013}.  Classifiers based on these model extensions could be derived in a similar fashion and compared to the other classification methods.   Also, we note that throughout this paper, we assume that the unobserved locations are missing completely at random (MCAR), meaning that the locations of the missing data are unrelated to other variables.  Of course, this may not be the case in practice, and future work could investigate the utility of these classifiers under other missing data scenarios.  

Finally, we note that in the empirical analysis in Section \ref{se:comparisons} we used a collapsed version of the original data due to computational considerations.  The original data consisted of 54,776 observations.  Although fitting any of the classification methods provided here to the full data set will be computational challenging, we note that low-rank methods such as those discussed within the paper \citep{reic:etal:2006, hugh:hara:2013, hank:etal:2015} can help ease the computational burden of fitting the SGLMM.

\section*{Acknowledgements}
The authors gratefully acknowledge support for this work from NASA's Land-Cover/Land-Use Change Program (NNG06GD31G), as well as from the National Science Foundation (DMS-0934595 and DMS-1209161).  They also thank Dr. Ningchuan Xiao, Department of Geography at The Ohio State University, for his assistance working with the land cover data, and the three anonymous referees for their helpful comments and suggestions for improving this paper.

\bibliography{references}

\newpage


\section*{APPENDICES} 

\begin{appendix}

\section{Other Classification Methods}\label{se:altclass}

In this appendix, we provide decision functions for the other classification methods discussed in Section \ref{se:spaceclass}.  For these alternative methods, we assume that the $\bfx$s only include the inputs, and thus do not include a term to allow for an intercept as in the model-based (GLM/GLMM) classification methods.  Unless otherwise noted, the descriptions of the classification methods are based on \cite{hast:etal:2001}.

\subsection{Model-based Non-spatial Classification}\label{se:nonspatialclass}

In this section we consider the corresponding non-spatial versions to the model-based spatial classification methods derived in Section \ref{se:spaceclass}.  For non-spatially dependent binary data, the underlying model is defined by equations (\ref{eq:randcomp})--(\ref{eq:systcomp}), where $\nu_i \equiv 0$ for all $i$.  In this case, $\bfomega = (\bfx^0, \bfbeta)$, 
\[
p_1(\bfomega) \equiv p_1(\bfx^0, \bfbeta) = g\inv(\bfx^{0\prime}\bfbeta),
\]
and $p_0(\bfomega) = 1-p_1(\bfomega)$.  It follows that the decision function is 
\begin{align}\label{eq:glmdecision1}
\delta_{GLM}(\bfomega) \equiv \delta_{GLM}(\bfx^0, \bfbeta) = \frac{p_1(\bfx^0, \bfbeta)}{p_0(\bfx^0, \bfbeta)}
= \frac{g\inv(\bfx^{0\prime}\bfbeta)}{1 - g\inv(\bfx^{0\prime}\bfbeta)}.
\end{align}
For the logit link function, 
\begin{equation}\label{eq:logitboundary}
\delta_{GLM-L}(\bfomega) = \delta_{GLM-L}(\bfx^0, \bfbeta)  = \exp\{\bfx^{0\prime}\bfbeta\},
\end{equation}
and for the probit link function,
\begin{equation}\label{eq:probitboundary}
\delta_{GLM-P}(\bfomega) = \delta_{GLM-P}(\bfx^0, \bfbeta) =  \frac{\Phi(\bfx^{0\prime}\bfbeta)}{1 - \Phi(\bfx^{0\prime}\bfbeta)}.
\end{equation}

For both decision functions, we can obtain maximum likelihood, posterior mean, and posterior predictive classifiers.  In practice, we evaluate the integral in (\ref{eq:postpredclass}) via Monte Carlo integration:  we approximate $\bar{p}_1(\bfomega)$ by drawing a realization of $Y^{0[t]} \sim \mbox{Bernoulli}(p_1(\bfx^0, \bfbeta^{[t]}))$ for $t=1, \dots, T$ and setting $\bar{p}_1(\bfomega) = \sum_{t=1}^T I(Y^{0[t]} =1)/T$, where $\bfbeta^{[t]}$ are the draws from the posterior distribution of $\bfbeta$,  and $T$ is the number of draws from the posterior distribution.

\subsection{Discriminant Analysis}\label{se:lda}

In discriminant analysis, rather than considering the explanatory variables $\bfx$ as fixed as they are in regression analyses, the $\bfx$s are viewed as random variables, with class-specific density functions $f_j(\bfx)$ corresponding to each class $C_j$.  The classes also have prior probabilities $\pi_j$, such that $\pi_0 + \pi_1 = 1$.  To determine the probability that a set of inputs will fall into class $j$, we employ Bayes' theorem which implies that
\begin{equation}\label{eq:bayes}
P(C_j | \bfx) = \frac{f_j(\bfx)\pi_j}{f_1(\bfx)\pi_1 + f_0(\bfx)\pi_0}.
\end{equation}
As with the model-based classification methods, the Bayes' classifier is to classify an observation to class $C_1$ when $P(C_1 | \bfx) > P(C_0 | \bfx)$ and to $C_0$ otherwise.

We first describe discriminant analysis in its general form, allowing an arbitrary form for the $f_j(\bfx)$s and the decision function, and then discuss threes special cases that we use in our analysis.  Let 
\[
\bfX_j = \left[ \begin{array}{c} \bfx_1\\ \vdots\\ \bfx_{n_j} \end{array} \right]
\]
where $\{\bfx_1, \dots, \bfx_{n_j}\} = \{\bfx_i: y_i = j\}$ and $\bfX_j$ is an $n_j \ell \times 1$ vector of inputs corresponding to observations in class $C_j$.  To classify $Y^0$, for each class we define
\[
\bfX_j^0 = \left[ \begin{array}{c} \bfx^{0}\\ \bfX_j \end{array}\right],
\]
where $\bfX_j^0$ is an $(n_j + 1)\ell \times 1$ vector and $\bfx^0$ is an $\ell\times 1$ vector of inputs associated with $Y^0$.  Our goal is to determine a decision boundary for classifying $Y^0$.

In discriminant analysis, $f_j(\cdot)$ is typically the multivariate normal density function,  
\[
f_j(\bfX_j^0) = \frac{1}{(2\pi)^{(n_j + 1)k/2} |\bfSigma_j^{X}|^{1/2}}\exp\left\{-\frac{1}{2}(\bfX_j^0 - \bfmu_j^{X})'\left(\bfSigma_j^{X}\right)\inv(\bfX_j^0 - \bfmu_j^{X})\right\},
\]
where $\bfmu_j^X$ is the $(n_j + 1) \ell \times 1$ class-specific mean vector and $\bfSigma_j^X$ is the $(n_j+1) \ell \times (n_j +1) \ell$ class-specific covariance matrix.  It follows that
\begin{equation}\label{eq:fjxo}
f_j(\bfx^0 | \bfx_1, \dots, \bfx_{n_j}) = \frac{1}{(2\pi)^{k/2} |\bfSigma_j^{x^0}|^{1/2}}\exp\left\{-\frac{1}{2}(\bfx^0 - \bfmu_j^{x^0})'\left(\bfSigma_j^{x^0}\right)\inv(\bfx^0 - \bfmu_j^{x^0})\right\}
\end{equation}
where 
\begin{align*}
\bfmu_j^{x^0} &= \bfmu_{j(\{1:\ell\})} + \bfSigma_{j(\{1:\ell\},\smneg\{1:\ell\})}^{X} \left(\bfSigma_{j (\smneg\{1:\ell\},\smneg\{1:\ell\})}^{X}\right)\inv (\bfX_j - \bfmu_{j(\smneg\{1:\ell\})})\\
\bfSigma_j^{x^0} &= \bfSigma_{j (\{1:\ell\},\{1:\ell\})}^{X} - \bfSigma_{j(\{1:\ell\},\smneg\{1:\ell\})}^{X} \left(\bfSigma_{j (\smneg\{1:\ell\},\smneg\{1:\ell\})}^{X}\right)\inv\bfSigma_{j(\smneg\{1:\ell\},\{1:\ell\})}^{X}.
\end{align*}
We use the subscript notation $(\{1:\ell\})$ to indicate the first $\ell$ elements of the corresponding matrix or vector (i.e., those indices corresponding to $\bfx^0$) and $(\smneg\{1:\ell\})$ indicates the matrix or vector without the first $\ell$ elements (i.e., the remaining indices corresponding to $\bfX_j$). 

Considering the log-odds, it follows from (\ref{eq:bayes}) and (\ref{eq:fjxo}) that
\begin{align}\label{eq:generallogodds}\begin{split}
 \log \left(\frac{P(C_1|\bfx^0)}{P(C_0|\bfx^0)}\right) 
= & \underbrace{\log \frac{\pi_1}{\pi_0} + \frac{1}{2}\log \frac{|\bfSigma_0^{x^0}|}{|\bfSigma_1^{x^0}|} -\frac{1}{2} \bfmu_1^{x^0\prime}\left(\bfSigma_1^{x^0}\right)\inv\bfmu_1^{x^0} + \frac{1}{2} \bfmu_0^{x^0\prime}\left(\bfSigma_0^{x^0}\right)\inv \bfmu_0^{x^0}}_{\equiv \alpha_0}\\
& + \bfx^{0\prime}\underbrace{\left((\bfSigma_1^{x^0})\inv\bfmu_1 - (\bfSigma_0^{x^0})\inv\bfmu_0^{x^0}\right)}_{\equiv \bfalpha_1} -  \bfx^{0\prime}\underbrace{\frac{1}{2}\left( (\bfSigma_1^{x^0})\inv - (\bfSigma_0^{x^0})\inv\right)}_{\equiv \bfalpha_2}\bfx^0.
\end{split}
\end{align}
Here, $\alpha_0$ is a scalar, $\bfalpha_1$ is an $\ell \times 1$ vector, and $\bfalpha_2$ is an $\ell \times \ell$ matrix, which we define for notational convenience.   A \emph{discriminant-analysis decision function} corresponding to the classification rule defined in equation (\ref{eq:class}) is 
\begin{align}
\delta_{DA}(\bfomega) \equiv &\, \delta_{DA}(\bfx^0, \bfmu_0^{x^0}, \bfmu_1^{x^0}, \bfSigma_0^{x^0}, \bfSigma_1^{x^0}) = \exp\{\alpha_0 + \bfx^{0\prime}\bfalpha_1 - \bfx^{0\prime}\bfalpha_2\bfx^0\}. \label{eq:generalDArule}
\end{align}

We now consider special cases of (\ref{eq:generalDArule}).  Each of these special cases assumes that the mean of $\bfx_i$ is equal across all observations, so that $\bfmu_j^{X} = [\bfone_{n_j + 1} \otimes \bfmu_j]$ where $\bfone_{n_j +1}$ is an $ (n_j +1) \times 1$ vector of ones and $\bfmu_j$ is an $\ell \times 1$ class specific mean vector.  The difference between each of these special cases is in the specification of the covariance matrix $\bfSigma_j^{X}$. We describe three popular discriminant analysis methods (linear discriminant analysis, diagonal linear discriminant analysis, and quadratic discriminant analysis) all of which assume that the $\bfx_i$ are independent.

Assuming the $\bfx_i$ are independent results in the following form for the covariance of $\bfX_j^0$:
\begin{equation}\label{eq:indepxivar}
\mbox{var}(\bfX_j^0) = \bfSigma_j^{X} = (\bfI_{n_j + 1} \otimes \bfLambda_j),
\end{equation}
where $\bfI_{n_j + 1}$ is an $(n_j + 1) \times (n_j +1)$ identity matrix and $\bfLambda_j$ is a class-specific covariance matrix for the $\ell$ components of $\bfx_i$.  Under this assumption, $\bfx^0$ is independent of $\bfx_1, \dots, \bfx_{n_j}$, so
\[
f_j(\bfx^0 | \bfx_1, \dots, \bfx_{n_j}) = f_j(\bfx^0) = \frac{1}{(2\pi)^{k/2} |\bfLambda_j|^{1/2}}\exp\{-\frac{1}{2}(\bfx^0 - \bfmu_j)'\bfLambda_j\inv(\bfx^0 - \bfmu_j)\}.
\]

Assuming a constant variance across classes (i.e., $\bfLambda_j = \bfLambda$ for $j = 0,1$) results in linear discriminant analysis (LDA) because the decision boundary is linear in the $\bfx$'s.  The log odds in equation (\ref{eq:generallogodds}) can be written in this case as
\begin{align*}
 \log \left(\frac{P(C_1|\bfx^0)}{P(C_0|\bfx^0)}\right) = \underbrace{\log \frac{\pi_1}{\pi_0} - \frac{1}{2} (\bfmu_1 + \bfmu_0)'\bfLambda\inv(\bfmu_1 - \bfmu_0)}_{\alpha_0^{LDA}} + \bfx^{0\prime}\underbrace{\bfLambda\inv(\bfmu_1 - \bfmu_0)}_{\bfalpha_1^{LDA}},
\end{align*}
where $\alpha_0^{LDA}$ is a scalar and $\bfalpha_1^{LDA}$ is an $\ell \times 1$ vector, defined for notational convenience. Therefore, the decision function for LDA is
\begin{align}
\delta_{LDA}(\bfomega) \equiv \delta_{LDA}(\bfx^0, \bfmu_0, \bfmu_1, \bfLambda) &= \exp\{\alpha_0^{LDA} + \bfx^{0\prime}\bfalpha_1^{LDA}\}.\label{eq:LDAdecision}
\end{align}
Note that this decision function is effectively equivalent to the one based on  the logistic regression model in (\ref{eq:logitboundary}), however, in logistic regression, we assume the $\bfx$'s are fixed and thus make no distributional assumptions on $\bfx$ as in discriminant analysis. 

In practice, the parameters $\pi_j, \bfmu_j, \bfLambda$ (and thus $\alpha_0^{LDA}$ and $\bfalpha_1^{LDA}$) are unknown but can be estimated using maximum likelihood:
\begin{itemize}
\item $\hat{\pi}_j = n_j/n$,
\item $\hat{\bfmu}_j = \sum_{i: y_i = j} \bfx_i/n_j$,
\item $\hat{\bfLambda} = \sum_{j \in \{0,1\}} \sum_{i:y_i=j} (\bfx_i - \hat{\bfmu}_j)(\bfx_i - \hat{\bfmu}_j)'/(n - 2)$.
\end{itemize}

Diagonal linear discriminant analysis (DLDA) additionally assumes independence between the $k$ inputs so that $\mbox{var}(\bfx_i) = \bfLambda$ is a diagonal matrix.  The {\it DLDA-based decision function} is the same as (\ref{eq:LDAdecision}), but using a diagonal matrix $\bfLambda$.  The $m^{th}$ diagonal element of $\bfLambda$ is estimated by $\hat{\bfLambda}_{(m,m)} = \sum_{j \in \{0,1\}} \sum_{i:y_i=j} (x_{im} - \hat{\mu}_{jm})^2/(n-2)$ where $x_{im}$  and $\hat{\mu}_{jm}$ are the $m^{th}$ elements of $\bfx_i$ and $\bfmu_j$, respectively.

In LDA, we assume a constant covariance for $\bfx_i$ among the classes (i.e., $\bfLambda_j = \bfLambda$ for $j = 0,1$). On the other hand, quadratic discriminant analysis (QDA) allows for each class to have its own covariance.  The log odds now contains a quadratic term in the $\bfx$'s:
\begin{align*}
\log \left(\frac{P(C_1|\bfx^0)}{P(C_0|\bfx^0)}\right)
 = &\underbrace{\log \frac{\pi_1}{\pi_0} + \frac{1}{2}\log \frac{|\bfLambda_0|}{|\bfLambda_1|} -\frac{1}{2} \bfmu_1'\bfLambda_1\inv\bfmu_1 + \frac{1}{2} \bfmu_0'\bfLambda_0\inv \bfmu_0}_{\alpha_0^{QDA}}\\
& + \bfx^{0\prime}\underbrace{(\bfLambda_1\inv\bfmu_1 - \bfLambda_0\inv\bfmu_0)}_{\bfalpha_1^{QDA}} -  \bfx^{0\prime}\underbrace{\frac{1}{2}(\bfLambda_1\inv - \bfLambda_0\inv)}_{\bfalpha_2^{QDA}}\bfx^0.
\end{align*}
Here,  $\alpha_0^{QDA}$ is a scalar, $\bfalpha_1^{QDA}$ is an $\ell \times 1$ vector, and $\bfalpha_2^{QDA}$ is an $\ell \times \ell$ matrix, which are defined for notational convenience.  This results in the {\it QDA-based decision function}:
\[
\delta_{QDA}(\bfomega) \equiv \delta_{QDA}(\bfx^0, \bfmu_0, \bfmu_1, \bfLambda_0, \bfLambda_1) = \exp\{\alpha_0^{QDA}  + \bfx^{0\prime}\bfalpha_1^{QDA} + \bfx^{0\prime}\bfalpha_2^{QDA}\bfx^0\},
\]
where we estimate the parameters by taking
\begin{itemize}
\item $\hat{\pi}_j = n_j/n$,
\item $\hat{\bfmu}_j = \sum_{i: y_i = j} \bfx_i/n_j$,
\item $\hat{\bfSigma}_j = \sum_{i: y_i = j} (\bfx_i - \hat{\bfmu}_j)(\bfx_i - \hat{\bfmu}_j)' / (n_j - 1)$.
\end{itemize}

\subsection{Support Vector Machines}\label{se:svm}
First introduced by \cite{cort:vapn:1992}, the goal of support vector machines (SVM) is to determine a hyperplane in covariate space separating the classes in such a way that the margin between the two classes is maximized.  The margin is the minimum distance between the the inputs $\bfx_i$ of the two classes in the direction perpendicular to the hyperplane.  The resulting function determining this hyperplane is the decision function.  

In SVM, the classes are labeled as either $1$ or $-1$ (instead of 1 or 0, as before).  To accommodate this convention, we redefine observations $y^*_i = 2y_i - 1$, for $i=1, \dots, n$, so that $y^*_i \in \{-1, 1\}$.

Consider the decision function  
\begin{equation}\label{eq:svmfunc}
\delta_{SVM}(\bfomega) \equiv \delta_{SVM}(\bfx, \bfbeta, \beta_0) = \exp\{\bfx'\bfbeta + \beta_0\}.
\end{equation}
Given observations $\{y^*_i, \bfx_i\}$ for $i = 1, \dots, n$, maximizing the margin between the two classes and the hyperplane is equivalent to minimizing $||\bfbeta||$ subject to $y^*_i(\bfx_i'\bfbeta + \beta_0) \ge 1$ for all $i = 1, \dots, n$.  This problem can be represented as the following Lagrange optimization function
\begin{equation}\label{eq:svmlagrange}
\max_{\bfzeta} L = \max_{\bfzeta} \left(\sum_{i=1}^n \zeta_i - \frac{1}{2} \sum_{i = 1}^n \sum_{i^*=1}^n \zeta_i \zeta_{i^*} y^*_i y^*_{i^*} \bfx_i' \bfx_{i^*}\right),
\end{equation}
where $\bfzeta= (\zeta_1, \dots, \zeta_n)'$ are the Lagrangian multipliers, which are subject to the constraints that $\sum_{i=1}^n \zeta_i y^*_i = 0$ and $\zeta_i \ge 0$ for all $i$.  The $\bfx_i$ where $\zeta_i > 0$ are the support vectors and are the only vectors which influence the position of the hyperplane.  
We can write the relationship between $\bfzeta$ and $\bfbeta$ as
\[
\bfbeta = \sum_{i=1}^n \zeta_i y_i^* \bfx_i
\]
and the relationship between $\bfzeta$ and $\beta_0$ as
\[
\beta_0 = \frac{1}{n_{sv}} \sum_{i: \zeta_i > 0 } (\bfbeta\bfx_i - y_i^*)
\]
where $n_{sv}$ is the number of support vectors.

While the above hyperplane is linear, SVM can be extended to create nonlinear boundaries between the classes.  This extension can be acheived by transforming the inputs into a space where they can be separated linearly, and again find the separating hyperplane in this transformed covariate space.  We can use the Lagrange optimization function in (\ref{eq:svmlagrange}) with the transformed inputs $h(\bfx_i, \bfx_{j})$:
\begin{equation}\label{eq:svmlagrangekernel}
\max L = \max \left(\sum_{i=1}^n \zeta_i - \frac{1}{2} \sum_{i = 1}^n \sum_{j=1}^n \zeta_i \,\zeta_{j}\, y^*_i\, y^*_{j}\, h(\bfx_i, \bfx_{j})\right)
\end{equation}
subject to $\sum_{i=1}^n \zeta_i y^*_i = 0$ and $0 \le \zeta_i \le \lambda$ for all $i$, where $\lambda$ is a tuning parameter allowing for crossover among the two classes and $h(\cdot, \cdot)$ is a symmetric positive (semi-) definite function. In our data analysis, we consider the following three popular kernels:
\begin{itemize}
\item Linear: $h(\bfx_i, \bfx_j) = \bfx_i' \bfx_j $
\item d$^{th}$ Degree Polynomial: $h(\bfx_i, \bfx_j) = (1 + \bfx_i'\bfx_j)^d$
\item Radial: $h(\bfx_i, \bfx_j) = \exp\{-u||\bfx_i - \bfx_j||^2\}$, where $u$ is a fixed constant.
\end{itemize}  

Now, (\ref{eq:svmfunc}) can be written as 
\begin{equation}\label{eq:svmfunch}
\delta_{SVM}(\bfomega) \equiv \delta_{SVM}(\bfx, \bfzeta, \beta_0, \lambda) = \sum_{i = 1}^n \zeta_i \, y^*_i \, h(\bfx, \bfx_i) + \beta_0,
\end{equation}
And $\hat{\delta}_{SVM}(\bfx^0, \bfzeta, \beta_0, \lambda) = \sum_{i = 1}^n \hat{\zeta}_i\, y_i^* \, h(\bfx^0, \bfx_i) + \hat{\beta}_0$.  

When implementing this classification method, we use the \texttt{R} package \texttt{e1071} \citep{e1071}, to compute $\hat{\zeta}_i$ and $\hat{\beta}_0$ via a quadratic optimization function for a fixed value of $\lambda$.

\subsection{k-Nearest Neighbors}
\subsubsection{k-Nearest Neighbors in Covariate Space} \label{se:knn-cov}
The k-nearest neighbors (kNN) classification method makes no assumptions about an underlying model.  Using this method, for a point $\{Y^0, \bfx^0\}$, the closest $k$ points $\{\bfx_{(r)}, r=1, \dots, k\}$ to $\bfx^0$ are identified, and $Y^0$ is assigned to the most popular class among the $k$ neighbors, where ties are broken at random.  ``Distance" here is measured in covariate space, not geographic space, and could be defined using any valid distance metric.  In our implementation of the method, we use Euclidean distance so that 
\[
d_{(r)} = ||\bfx_{(r)} - \bfx^0||.
\]
Here, $d_{(i)}$ represents the ordered distances where the minimum is $d_{(1)}$ and the maximum is $d_{(n)}$, and $\bfx_{(r)}$ are the $\bfx_i$ corresponding to $d_{(r)}$.  Using this measure of distance requires standardization of the variables so that no variable is given more weight than another.  

For the binary case, a decision function can then be defined as
\begin{equation*}
\delta_{kNN-C}(\bfomega) \equiv \delta_{kNN-C}(\bfx^0, \bfx, \bfY) = 2\sum_{r=1}^k y_{(r)}/k,
\end{equation*}
where $y_{(r)}$ is the $y_i$ associated with $d_{(r)}$.
When classifying $Y^0$, if $\delta_{kNN-C}(\bfomega) = 1$, $y_{pred}^0 = 1$ with probability $.5$ and $y_{pred}^0 = 0$ with probability $.5$.

\subsubsection{k-Nearest Neighbors in Geographic Space} \label{se:knn-geo}

Instead of using covariates to determine proximity as in Section \ref{se:knn-cov}, we can also use geographic space.  Using this approach, for prediction at location $\bfs^0$, the closest $k$ observed points are identified $\{\bfs_{(r)}, r = 1, \dots, k\}$, and $Y^0$ is assigned to the most popular class among its $k$ geographic neighbors.  Just as in k-Nearest Neighbors for covariate space, we can use any distance metric, but we again use Euclidean distance so that
\[
d^*_{(r)} = ||\bfs_{(r)} - \bfs^0||.
\]
The $d_{(i)}^*$  represent ordered distances due to geographical space, and $\bfs_{(r)}$ are the $\bfs_i$ corresponding to $d_{(r)}^*$.  The decision function for kNN based on geographic space can then be defined as
\[
\delta_{kNN-G}(\bfomega) \equiv \delta_{kNN-G}(\bfs^0, \bfs, \bfY) = 2 \sum_{1}^k y^*_{(r)}/k,
\]
where $y^*_{(r)}$ are the $y_i$ associated with $d_{(r)}^*$.  When classifying $Y^0$, if $\delta_{kNN-G}(\bfomega) = 1$, we assign $y_{pred}^0 = 1$ with probability $.5$ and $y_{pred}^0 = 0$ with probability $.5$.

\subsection{Spatial Extensions of Discriminant Analysis}

\subsubsection{Switzer}

\cite{swit:1980} extend LDA by augmenting the covariates of the focal location with an average of neighboring covariate values.  Specifically, let $\bfx^{\star} = (\bfx^{0\prime}, \bfx^{c\prime})'$, where 
\[
\bfx^{c} = \frac{1}{n^\star} \sum_{j =1}^{n^\star} \bfx_{j},
\]
where $\bfx_{j}$ for $j = 1, \dots, n^\star$ are the covariate values of the $n^\star$ neighbors of the focal location.  In our empirical analysis, we use second-order neighborhood structure, where locations sharing an edge or corner with the focal location are considered neighbors.  Then, the decision function is the same as the decision function for LDA (Equation \ref{eq:LDAdecision}), where we replace $\bfx^{0}$ with  $\bfx^{\star}$.

\subsubsection{Mardia}\label{se:Mardia} 

\cite{mard:1984} uses the same idea as \cite{swit:1980}, but weights the augmented covariates according to a spatial correlation matrix.  In this case, for the focal location, we choose a ``window" of neighbors which defines $n^\star$ neighbors around the focal location.  Often, a $3 \times 3$ window is used (this corresponds to a second-order neighborhood structure), and we used this in our analysis.   We denote the $(n^\star + 1) \times \ell $ matrix of covariates belonging to the focal location and its $n^\star$ neighbors by 
\[
\bfX^{\star} = \left[\begin{array}{c}
\bfx^{0\prime}\\
\bfx_1' \\
\vdots \\
\bfx_{n^\star}'
\end{array}\right].
\]
The decision function can then be written as
\begin{align*}
\delta_{Mardia}(\bfomega) &\equiv \delta_{Mardia}(\bfX^\star, \bfmu_0, \bfmu_1, \bfLambda_0, \bfLambda_1, \bftheta_0, \bftheta_1) = \frac{S_1}{S_0}
\end{align*}
where
\begin{align*}
S_j &= \log \pi_j - \frac{n^\star+1}{2} \log |\bfLambda_j| + \ell(n^\star+1) \log \psi_j - \frac{1}{2 \psi_i^2} (G_j - \psi_j^2\bfmu_j)'\bfLambda_j\inv(G_j- \psi_j^2\bfmu_j),\\
G_j & = \bfone'K^0(\bftheta_j) \bfX^\star, \\
\psi_j^2 & = \bfone'K^0(\bftheta_j)\bfone,
\end{align*}  
and $\bfmu_j$ is the $\ell \times 1$ class-specific mean, $\bfLambda_j$ is the $\ell \times \ell$ class-specific covariance matrix among the covariates, $K^0(\bftheta_j)$ is the $(n^\star + 1) \times (n^\star + 1)$ spatial correlation matrix of the focal location and its $n^\star$ neighbors parameterized by the class-specific spatial covariance parameters, $\bftheta_j$. 

For a fixed spatial correlation matrix, \cite{mard:1984} provides a method to estimate the other parameters.  For the training data in class $j$, let
\[
\bfX_{(j)} = \left[ \begin{array}{c}
\bfx_1'\\
\vdots\\
\bfx_{n_j}'
\end{array} \right]
\]
be an $n_j \times \ell$ matrix of covariates $\{\bfx_1, \dots, \bfx_{n_j}\} = \{\bfx_i : y_i=j\}$.  Then, 
\begin{align*}
\hat{\bfmu}_j &= \frac{\bfone'K(\bftheta_j)\inv\bfX_{(j)}}{\bfone'K(\theta_j)\inv\bfone}\\
\hat{\bfLambda}_j & = \frac{1}{n_j}  \left(\bfX_{(j)} -  (\bfone \otimes \hat{\bfmu}_j')\right)' \,K(\bftheta_j)\inv \left(\bfX_{(j)} -  (\bfone \otimes \hat{\bfmu}_j')\right),
\end{align*}
where $K(\bftheta_j)$ is the fixed $n_j \times n_j$ spatial correlation matrix of class $j$.  To estimate the spatial correlation, they recommend using covariograms of the covariates in the training data for each class.  Finally, as with LDA, we can also assume the same covariance and spatial correlation structure across the classes.  

\subsubsection{Spatial LDA} 
\cite{salt:duci:2005} define a spatio-temporal discriminant analysis approach.  Because the data we model is not temporal,  we provide a space-only version of their model here.
\[
\delta_{SLDA}(\bfomega) \equiv \delta_{SLDA}(\bfX, \bfu^0, \bfB_1, \bfB_0, \bfLambda) = \left( \bfx^0 - \frac{{\bfmu}_1 + {\bfmu}_0 }{2}\right)' {\bfLambda}\inv ({\bfmu}_1 - {\bfmu}_0) + \log\left(\frac{\pi_1}{\pi_0}\right)
\]
where $\bfx^0$ are the inputs at the location of interest, $\bfLambda$ is the covariance among the covariates, $\bfmu_j$ is the class-specific mean modeled with a $q \times 1$ vector of regressors, $\bfu^0$, such that
\[
\bfmu_j = \bfB_j'\bfu^0,
\]
and $\bfB_j$ is a $q \times \ell$ class-specific matrix of coefficients.  Often, $\bfu^0$ is just the location coordinates.  

To estimate these parameters, let 
\begin{align*}
\bfU_j = \left[\begin{array}{c} \bfu_1'\\ \vdots \\ \bfu_{n_j}'\end{array}\right].
\end{align*} 
Then, for a fixed spatial correlation matrix, $K(\bftheta_j)$, the maximum likelihood estimators are
\begin{align*}
\hat{\bfB}_j &= (\bfU_j'K(\bftheta_j)\inv\bfU_j)\inv\bfU_j'K(\bftheta_j)\inv\bfX_{(j)},\\
\hat{\bfSigma} & = \frac{1}{n_1 + n_0} \sum_{j = 0}^1 (\bfX_{(j)} - \bfU_j\hat{\bfB}_j)'K(\bftheta_j)\inv(\bfX_{(j)} - \bfU_j\hat{\bfB}_j),
\end{align*}
where $\bfX_{(j)}$ is as defined in Section \ref{se:Mardia}. Again we use the variogram to estimate $\bftheta_j$.

\subsection{Press} 

The final spatial method from \cite{pres:1996} uses the class to determine neighborhoods for evaluating the probabilities of the focal point belonging to each class.  These probabilities are found by marginalizing over the parameters in the likelihood distribution.  Their algorithm is as follows:
\begin{enumerate}
\item Using the training data, draw samples from the posterior distributions of the parameters for each class.  In our analysis, for class $j$, we used a multivariate normal distribution allowing for spatial dependence among the covariates.  Specifically, we assumed 
\[
\bfX_{j} \sim \mathcal{N}(\bfone \otimes \bfmu_j, K(\bftheta_j) \otimes \bfLambda_j),
\]
where $\bfX_j$ is the $\ell n_j \times 1$ vector of covariates in class $j$, $\bfmu_j$ is the $\ell \times 1$ class-specific mean, $K(\bftheta_j) \equiv K(\theta_j)$ is a spatial correlation matrix, and $\bfLambda_j$ is the $\ell \times \ell$ class-specific covariance among the covariates.  We used the following prior distributions:
\begin{align*}
 \bfmu_j & \sim \mathcal{N}(\bfzero, 10000\,\bfI)\\
 \theta_j &\sim \mbox{U}(0, 20)\\
\bfLambda_j &\sim \mbox{InvWishart}(5, \bfI).
\end{align*}
\item Pre-classify each location using a ``zero-neighbor Bayesian" classification.  Using the samples from the posterior distributions, we computed
\begin{align*}
p_{j}^{0[t]} &= f(\bfx^0 | \bfmu_j^{[t]}, \bfLambda_j^{[t]})\\
Y^{0[t]} &= \begin{cases}
1, \mbox{ if } p_{1}^{0[t]} > p_{0}^{0[t]} \\
0, \mbox{ otherwise}
\end{cases}
\end{align*}
where $f(\cdot|\bfmu_j^{[t]}, \bfLambda_j^{[t]})$ is the normal distribution with the $t$th draw of $\bfmu_j$ as the mean, and the $t$th draw of $\bfLambda_j$ as the covariance.  $Y^0$ is then pre-classified to the most frequent class among the $Y^{0[t]}$.  Denote the set of all pre-classified observations by $\bfy^{pc} = \{y_1^{pc}, \dots, y_n^{pc}\} $.
\item For each location, identify a homogoneous neighborhood using a degree of homogeneity measure to compare possible neighborhoods.  In our analysis, we compared four possible directional neighborhoods -- north, south, east, and west -- where the neighborhood made up the 8 locations in the named direction along with the focal location. The neighborhood for class $j$ is the neighborhood  with the most locations pre-classified to class $j$.  If there were multiple neighborhoods with the same number of locations in class $j$, we randomly selected one.  Call the selected neighborhood for class $j$ of the focal location $N_{j}^0$. 
\item For each location,  compute 
\[
\Delta^0_{j} \propto \pi_j P(\bfX^0_{j}|y^{0pc} = j, y_1^{pc} = j, \dots y^{pc}_{n_j\star} = j), 
\]
where $y^{0pc}$ is the pre-classified value of the focal location,  $y_1^{pc}, \dots, y^{pc}_{n_j\star}$ correspond to the $n_j^\star$ neighbors pre-classified to $j$ in $N_{j}^0$ and 
\[
\bfX^0_{j} = \left[ \begin{array}{c}
\bfx^0\\
\bfx_1 \\
\vdots \\
\bfx_{n_j^\star}
\end{array}\right].
\]
In our analysis, we computed
\begin{align*}
\Delta_{j}^{0[t]} &= \pi_j f(\bfX^0_{j} | \bfmu_j^{[t]}, \theta_j^{[t]}, \bfLambda_j^{[t]})\\
\hat{Y}^{0[t]} & = \begin{cases}
1, \mbox{ if } \Delta_{i1}^{[t]} > \Delta_{i0}^{[t]} \\
0, \mbox{ otherwise}
\end{cases}
\end{align*}
where $f(\cdot|  \bfmu_j^{[t]}, \theta_j^{[t]}, \bfLambda_j^{[t]}) = \mathcal{N}(\bfone \otimes  \bfmu_j^{[t]}, K(\theta_j^{[t]}) \otimes \bfLambda_j^{[t]})$.  Finally, $y^0_{pred}$ is classified to be the most frequent class among the $\hat{Y}^{0[t]}$.  \end{enumerate}

We can write the decision function as
\begin{align*}
\delta_{Press}(\bfomega) \equiv \delta_{Press}(\bfX^0_0, \bfX^0_1, \bfmu_0, \bfmu_1, \bftheta_0, \bftheta_1, \bfLambda_0, \bfLambda_1) = \frac{\Delta_1^0}{\Delta_0^0},
\end{align*}
where $\Delta_j^0 =  \pi_j f(\bfX^0_{j} | \bfmu_j, \theta_j, \bfLambda_j)$.  Note that the algorithm described above corresponds to the posterior predictive classifier (Section \ref{se:classprob}) based on this decision function because it marginalizes over the posterior distributions of parameters $\bfmu_j$, $\theta_j$, and $\bfLambda_j$.

For more details and options for other approaches to pre-classification and selecting a directional neighborhood, we refer the reader to \cite{pres:1996}.  

\section{Model-fitting Algorithms for the SGLM/SGLMM}\label{se:modest}

\subsection{Data Augmentation Algorithms}\label{se:DA}

In Section \ref{se:SGLMs}, $\epsilon_i$ in equation (\ref{eq:ACSGLMM}) has a fixed variance of 1.  This is for identifiability purposes and also creates the probit link.  However, for model-fitting purposes, instead of fixing the variance parameter to $1$, we can use and marginalize over a variance parameter \citep[for examples in the Bayesian probit model, see][]{imai:vand:2005, berr:cald:2012}.   \cite{berr:cald:2012} describe a model-fitting algorithm for the probit SGLM.  Here we show how this algorithm can be extended to create a data augmentation model-fitting algorithm for the SGLMM.  We briefly review the model notation and model-fitting algorithm here and then provide an adapted algorithm for the probit SGLMM.  In both algorithms, we make use of a non-identifiable variance parameter of the latent variable to facilitate mixing of the MCMC.  Within this appendix, we use the $\tilde{ }$ notation to represent unidentifiable parameters.  For both algorithms, we used what \cite{berr:cald:2012} call the Non-collapsed Marginal-Scheme 1 Algorithm.

Equation (\ref{eq:spatdep}) in Section \ref{se:SGLMs} describes the spatial covariance matrix $\bfSigma(\bftheta)$ to be a scalar, $\theta_1$, times a spatial dependence matrix $K(\bftheta_2)$.  For the SGLM, $\theta_1$ is the non-identifiable  parameter which \cite{berr:cald:2012} make use of as a working parameter.  For consistency of the $\tilde{ }$ notation, let $\tilde{K}(\bftheta) = \sigma^2 K(\bftheta)$, where $\tilde{K}(\bftheta) = \bfSigma(\bftheta)$ is the spatial covariance matrix, $\sigma^2 = \theta_1$ is a non-identifiable scalar used as the working parameter of the algorithm, and $K(\theta)$ is the identifiable spatial dependence matrix parameterized by a single parameter, $\theta$.  

The data augmentation model-fitting algorithm for the SGLM is then

\begin{tabular}{lll}
& \emph{Step 1:} &Sample $\sigma^2_{temp} \sim \pi(\sigma^2)$\\
&&Sample $\tilde{\bfZ} | \bfY, \bfbeta, \theta, \sigma^2_{temp}$\\
&&Set $\bfZ = \tilde{\bfZ}/\sigma_{temp}$\\
& \emph{Step 2:} & Sample $(\sigma^2, \bfbeta)| \tilde{\bfZ}, \bfY, \theta$\\
&& Set $\bfbeta = \tilde{\bfbeta}/\sigma$\\
&\emph{Step 3:} & Sample $\theta | \tilde{\bfZ}, \bfY, \tilde{\bfbeta}, \sigma^2$.
\end{tabular}

For more details, see \cite{berr:cald:2012}.

For the SGLMM,  we can build on this model-fitting algorithm by adding an additional step to sample the additional parameter.  Recalling that $\epsilon_i \sim \mathcal{N}(0, 1)$, we can write 
\[
(\bfnu + \tilde{\bfepsilon}) \sim \mathcal{N}(\bfzero, \tilde{\bfSigma}^*(\bftheta)),
\]
where 
\[
\tilde{\bfSigma}^*(\bftheta) = \bfI + \bfSigma(\bftheta) = \bfI + \sigma^2K(\theta),
\]
where $\bfSigma(\theta)$ is defined as in the SGLM algorithm above.  Letting $\gamma^2 = 1 + \sigma^2$ and $\kappa = \sigma^2/\gamma^2$, we obtain
\[
\tilde{\bfSigma}^*(\bftheta) = \gamma^2\left( \,(1 - \kappa) \bfI\, +\, \kappa\, K(\bftheta_2)\, \right).
\]
This is simply a scalar times a spatial dependence matrix, just as in the model-fitting algorithm for the SGLM.  Notice that $\kappa$ is bounded by 0 and 1 and that if $\kappa = 1$, $\tilde{\bfSigma}^*(\bftheta) = \sigma^2 K(\theta)$, or the spatial covariance matrix of the SGLM.  Therefore, to adapt the SGLM data augmentation algorithm to the SGLMM, we let $\gamma^2$ be the non-identifiable working parameter and add a step to sample $\kappa$.

Here we provide the model-fitting algorithm and the full conditional distributions for the SGLMM.  We use priors $\bfbeta \sim \mathcal{N}(\bfzero, \bfV_\bfbeta)$, $\gamma^2 \sim b_\gamma(\chi^2_{a_\gamma})\inv$, $\theta \sim \pi(\theta)$, and $\kappa \sim \mbox{U}(0, 1)$.  We use superscript $[t]$ to denote the value of a parameter at the $t$th iteration of the algorithm.  Let $\tilde{\bfSigma}^*(\bftheta) = \gamma^2 \bfSigma^*(\bftheta)$, where $\bfSigma^*(\bftheta) = (1-\kappa)\bfI + \kappa K(\theta)$.

\begin{description}
\item[Step 1:]  Sample $\bfZ\cur$ from $\bfZ|\bfY, \bfbeta\prev, \theta\prev, \kappa\prev$:\\
 Draw $\gamma^2_{temp} \sim \pi(\gamma^2)$.\\
 For $i = 1, \dots, n$, define $Z\prev_{\neg i} = (Z\cur_1, \dots, Z\cur_{i-1}, Z\prev_{i+1}, \dots, Z_n\prev)'$ and sample $\tilde{Z}_i$ from 
\begin{align*}
\tilde{Z}_i|\bfY, \bfZ\prev_{\neg i}, \bfbeta\prev, \theta\prev, \kappa\prev, \gamma^2_{temp} \sim \begin{cases}
TN(\mu_{\tilde{Z}_i}, \tau^2_{\tilde{Z}_i}, 0, \infty), &\mbox{if  } Y_i = 1\\
TN(\mu_{\tilde{Z}_i}, \tau^2_{\tilde{Z}_i}, -\infty, 0), &\mbox{if  } Y_i = 0
\end{cases},
\end{align*}
where $TN(\mu_{\tilde{Z}_i}, \tau^2_{\tilde{Z}_i}, \ell, u)$ is a truncated normal distribution with lower and upper bounds $\ell$ and $u$, respectively, and mean and variance 
\begin{align*}
\mu_{\tilde{Z}_i}& = \gamma_{temp}\bfx_i'\bfbeta\prev + [\bfSigma^*(\bftheta\cur)]_{i,\neg i} \left( [\bfSigma^*(\bftheta\cur)]_{\neg i,\neg i} \right)\inv \gamma_{temp}\left( \bfZ\prev_{\neg i} - \bfX_{\neg i} \bfbeta\prev\right)\\
 \tau^2_{\tilde{Z}_i} & = \gamma^2_{temp}\left([\bfSigma^*(\bftheta\cur)]_{i, i} - [\bfSigma^*(\bftheta\cur)]_{i,\neg i} \left([\bfSigma^*(\bftheta\cur)]_{\neg i,\neg i}\right)\inv [\bfSigma^*(\bftheta\cur)]_{\neg i, i} \right).
\end{align*}
Set $Z\cur_i = \tilde{Z}\cur_i/\gamma_{temp}$.

For an unobserved location, sample
\[
\tilde{Z}^{0}| \bfx^0, \bfY, \bfZ\cur, \bfbeta\prev, \theta\prev, \kappa\prev, \gamma^2_{temp} \sim \mathcal{N}( \gamma_{temp}\mu_{Z^0}, \gamma^2_{temp}\sigma^2_{Z^0}), 
\]
where $\mu_{Z^0}$ and $\sigma^2_{Z^0}$ are defined in equations (\ref{eq:predmean}) and (\ref{eq:predvar}), and set $Z^{0[t]} = \tilde{Z}^0/\gamma_{temp}$.

\item[Step 2:] Sample $(\gamma^2)\cur, \bfbeta\cur$ from $\gamma^2, \bfbeta | \bfY, \tilde{\bfZ}\cur, \theta\prev, \kappa\prev$:\\
Sample
\[
(\gamma^2)\cur \sim \left( (\tilde{\bfZ} - \bfX\hat{\bfbeta})'\left(\bfSigma^*(\bftheta\prev)\right)\inv (\tilde{\bfZ} - \bfX\hat{\bfbeta}) + b_\gamma^2 + \hat{\bfbeta}'\bfV_\bfbeta\inv\hat{\bfbeta}\right) (\chi^2_{n+a_\gamma})\inv,
\] where $\hat{\bfbeta} = \left(\bfX'\left(\bfSigma^*(\bftheta\prev)\right)\inv\bfX + \bfV_\bfbeta\inv\right)\inv\bfX'\left(\bfSigma^*(\bftheta\prev)\right)\inv\tilde{\bfZ}\cur$.
Sample
\[
\tilde{\bfbeta} \sim \mathcal{N}\left(\hat{\bfbeta}, (\gamma^2)\cur\left(\bfX\left(\bfSigma^*(\bftheta\prev)\right)\inv\bfX + \bfV_\bfbeta\inv\right)\inv\right).
\]
Set $\bfbeta\cur = \tilde{\bfbeta}/\gamma\cur$.
\item[Step 3:] Sample $\theta\cur$ from $\theta|\bfY, \tilde{\bfZ}\cur, \tilde{\bfbeta}\cur, (\gamma^2)\cur, \kappa\prev$ via a random walk Metropolis step:\\
Sample a proposal value ${\theta_{prop}}$ from a proposal distribution.  We used a normal distribution, $\mathcal{N}(\theta|\theta\prev, \tau^2_\theta)$, where $\tau^2_\theta$ is the fixed variance of the proposal distribution.  Define
\[
\theta\cur = \begin{cases}
\theta_{prop} & \mbox{with probability } c(\theta\prev, \theta_{prop})\\
\theta\prev & \mbox{with probability } 1 - c(\theta\prev, \theta_{prop})
\end{cases}
\]
where
\[
c(\theta\prev, \theta_{prop}) = \min\left\{ \frac{\pi(\theta_{prop}|\bfY, \tilde{\bfZ}, \tilde{\bfbeta}, (\gamma^2)\cur, \kappa\prev)}{\pi(\theta\prev|\bfY, \tilde{\bfZ}, \tilde{\bfbeta}, (\gamma^2)\cur, \kappa\prev)}, 1\right\}.
\]
The posterior distribution of $\theta$ in the acceptance probability is
\[
\pi(\theta_{prop}|\bfY, \tilde{\bfZ}, \tilde{\bfbeta}, (\gamma^2)\cur, \kappa\prev) \propto \phi(\tilde{\bfZ}; \bfX\tilde{\bfbeta}, \gamma^{2[t]}\bfSigma^*(\bftheta))\pi(\theta),
\]
where $\phi(\cdot)$ is the multivariate normal density function.  
\item[Step 4:]Sample $\kappa\cur$ from $\theta|\bfY, \tilde{\bfZ}\cur, \tilde{\bfbeta}\cur, (\gamma^2)\cur, \theta\cur$ via a random walk Metropolis step:\\
Sample a proposal value ${\kappa_{prop}}$ from a proposal distribution.  We used a normal distribution, $\mathcal{N}(\kappa|\kappa\prev, \tau^2_\kappa)$, where $\tau^2_\kappa$ is the fixed variance of the proposal distribution.  Define
\[
\kappa\cur = \begin{cases}
\kappa_{prop} & \mbox{with probability } c(\kappa\prev, \kappa_{prop})\\
\kappa\prev & \mbox{with probability } 1 - c(\kappa\prev, \kappa_{prop})
\end{cases}
\]
where
\[
c(\kappa\prev, \kappa_{prop}) = \min\left\{ \frac{\pi(\kappa_{prop}|\bfY, \tilde{\bfZ}, \tilde{\bfbeta}, (\gamma^2)\cur, \theta\cur)}{\pi(\kappa\prev|\bfY, \tilde{\bfZ}, \tilde{\bfbeta}, (\gamma^2)\cur, \theta\cur)}, 1\right\}.
\]
The posterior distribution of $\theta$ in the acceptance probability is
\[
\pi(\kappa_{prop}|\bfY, \tilde{\bfZ}, \tilde{\bfbeta}, (\gamma^2)\cur, \theta\cur) \propto \phi(\tilde{\bfZ}; \bfX\tilde{\bfbeta}, \gamma^2\bfSigma^*(\bftheta))\pi(\kappa),
\]
where $\phi(\cdot)$ is the multivariate normal density function.  

\end{description}

\subsection{Low-rank Probit SGLMM}\label{se:lowrank}

In this section, we provide an overview and a data augmentation algorithm for fitting the \cite{hugh:hara:2013} probit version of their low-rank SGLMM.  

Originally proposed by \cite{reic:etal:2006}, \citeauthor{hugh:hara:2013} decompose the $\nu_i$'s in equation (\ref{eq:systcomp}) into two pieces: one collinear with $\bfX$ (this component is discarded) and one orthogonal to $\bfX$.   \citeauthor{hugh:hara:2013} use the Moran operator to do this.  Letting $\bfA$ be the binary neighborhood matrix, where the $ij$th element is 1 if locations $i$ and $j$ are neighbors, and 0 otherwise, the Moran operator is defined to be 
\[
\bfM = \left(\bfI - \bfX\left(\bfX'\bfX\right)\inv \bfX'\right) \bfA \left(\bfI - \bfX\left(\bfX'\bfX\right)\inv \bfX'\right). 
\]
Let $\bfPsi$ be the $n \times r$ eigenvectors of $\bfM$ corresponding to the $r$ largest eigenvalues.  For gridded data, \citeauthor{hugh:hara:2013} propose letting $r$ be equal to 10 percent of the eigenvectors.  We can then adjust equation (\ref{eq:ACSGLMM}) to reflect this low rank version of the SGLMM: 
\[
Z_i = \bfx_i'\bfbeta_1 + \bfPsi_i\bfbeta_2 + \epsilon_i,
\]   
where $\bfbeta_1$ is the $\ell \times 1$ vector of coefficients corresponding to the $\ell$ covariates and $\bfbeta_2$ is the $r \times 1$ vector of coefficients corresponding to $\bfPsi_i$, the $i$th row of the the selected $r$ Moran operator eigenvectors.  

In this approach, the spatial dependence is captured as additional covariates and coefficients.  Letting 
\[
\bfX_H = \left[ \begin{matrix}
\bfX & 
\bfPsi
\end{matrix}\right]
\]
and
\[
\bfbeta_H = \left[ \begin{matrix}
\bfbeta_1\\
\bfbeta_2
\end{matrix}\right],
\]
we can write the spatial latent variable Bayesian probit model in the form of a non-spatial latent variable Bayesian probit model:
\[
Z_i = \bfx_{Hi}' \bfbeta_H + \epsilon_i,
\]where $\bfx_{Hi}$ is the $(\ell + r) \times 1$ vector corresponding to the $i$th row of $\bfX_H$ and  $\epsilon_i \sim \mathcal{N}(0, \gamma^2)$.  Setting prior distributions to be
\begin{align*}
\bfbeta_H \sim \mathcal{N}(\bfzero, \bfV_H),\\
\gamma^2 \sim b_\gamma (\chi^2_{a_\gamma})\inv,
\end{align*}
we use the following data augmentation model-fitting algorithm:
\begin{description}
\item[Step 1:] Sample $\bfZ^{[t]}$ from $\bfZ|\bfY, \bfbeta_H^{[t-1]}$:\\
Draw $\gamma^2_{temp} \sim \pi(\gamma^2)$.\\
For $i = 1, \dots, n$, sample $\tilde{Z}_i$ from
\[
\tilde{Z}_i | \bfY, \bfbeta_H^{[t-1]} \sim 
\begin{cases} 
TN(\gamma_{temp}\, \bfx_{Hi}'\,\bfbeta_H, \gamma^2_{temp}, 0, \infty), & \mbox{ if } Y_i = 1\\
TN(\gamma_{temp}\,\bfx_{Hi}'\,\bfbeta_H, \gamma^2_{temp},  -\infty, 0), & \mbox{ if } Y_i = 0\\
\end{cases}.
\] 
Set $Z_i^{[t]} = \tilde{Z}_i/\gamma_{temp}$.

For an unobserved location, sample $\tilde{Z}^0$ from $\tilde{Z}^0 | \bfbeta_H^{[t-1]} \sim \mathcal{N}(\gamma_{temp}\, \bfx_H^{0\prime}\,\bfbeta_H, \gamma^2_{temp})$ and set $Z^{0[t]} = \tilde{Z}^0/\gamma_{temp}$.
\item[Step 2:] Sample $(\gamma^2)^{[t]}, \bfbeta_H^{[t]}$ from $\gamma^2, \bfbeta_H|\tilde{\bfZ}, \bfY$:

Sample
\[
(\gamma^2)\cur \sim \left( (\tilde{\bfZ} - \bfX_H\hat{\bfbeta}_H)'(\tilde{\bfZ} - \bfX_H\hat{\bfbeta}_H) + b_\gamma^2 + \hat{\bfbeta}'_H\bfV_H\inv\hat{\bfbeta}_H\right) (\chi^2_{n+a_\gamma})\inv,
\] where $\hat{\bfbeta}_H = \left(\bfX_H'\bfX_H + \bfV_H\inv\right)\inv\bfX_H'\tilde{\bfZ}\cur$.\\
Sample
\[
\tilde{\bfbeta}_H \sim \mathcal{N}\left(\hat{\bfbeta}_H, (\gamma^2)\cur\left(\bfX_H\bfX_H + \bfV_H\inv\right)\inv\right).
\]
Set $\bfbeta_H\cur = \tilde{\bfbeta}_H/\gamma\cur$.
\end{description}

\section{Simulation Study Classification Error Rates}\label{se:simrates}
Error rates for the simulation study in Section \ref{se:robust} are provided in the tables below.

\begin{tabular}{|c|c|cc|cc|cc|}
\hline
\multicolumn{8}{|l|}{ \textbf{ One-at-a-Time Training }}\\ \hline
\multicolumn{2}{|c|}{ $\kappa$ }& \multicolumn{2}{c|}{0.25} & \multicolumn{2}{c|}{0.5} & \multicolumn{2}{c|}{1}\\
 \multicolumn{2}{|c|}{ Model Fit }& SGLM & SGLMM & SGLM & SGLMM &SGLM & SGLMM \\ \hline 
Linear Component & Data Set &&&&&&\\ \cline{1-2}
Intercept & 1 & 0.3859 & 0.3859 & 0.4048 & 0.4118 & 0.2177 & 0.2381 \\ 
& 2 & 0.3232 & 0.3502 & 0.3795 & 0.3861 & 0.2287 & 0.2253 \\ 
& 3 & 0.3432 & 0.3432 & 0.2466 & 0.274 & 0.2905 & 0.3041 \\ 
\hline
Simple-1 & 1 & 0.3345 & 0.3412 & 0.2774 & 0.2774 & 0.1347 & 0.1481 \\ 
& 2 & 0.3255 & 0.3255 & 0.3028 & 0.3134 & 0.1259 & 0.1399 \\ 
& 3 & 0.354 & 0.354 & 0.3428 & 0.3498 & 0.1443 & 0.1512 \\ 
\hline
Simple-2 & 1 & 0.3559 & 0.3593 & 0.2891 & 0.2993 & 0.1103 & 0.1345 \\ 
& 2 & 0.323 & 0.3127 & 0.2432 & 0.25 & 0.1424 & 0.1458 \\ 
& 3 & 0.3017 & 0.3051 & 0.3389 & 0.3423 & 0.1577 & 0.1846 \\ 
\hline
Multiple & 1 & 0.3172 & 0.331 & 0.2449 & 0.2585 & 0.1513 & 0.1612 \\ 
& 2 & 0.2902 & 0.3217 & 0.2517 & 0.2586 & 0.1533 & 0.1672 \\ 
& 3 & 0.2437 & 0.2616 & 0.2457 & 0.2803 & 0.1325 & 0.1358 \\ 
\hline
Confounded & 1 & 0.2399 & 0.2601 & 0.2068 & 0.2271 & 0.0993&0.0925 \\ 
& 2 & 0.2295 & 0.2226 & 0.2990 &0.3156 & 0.0976 & 0.1077\\ 
& 3 &  0.2748 & 0.2682 & 0.2517 & 0.2653 & 0.1559 & 0.1661 \\ 
\hline
\end{tabular}

\begin{tabular}{|c|c|cc|cc|cc|}
\hline
\multicolumn{8}{|l|}{ \textbf{ Joint Training }}\\ \hline
\multicolumn{2}{|c|}{ $\kappa$ }& \multicolumn{2}{c|}{0.25} & \multicolumn{2}{c|}{0.5} & \multicolumn{2}{c|}{1}\\
 \multicolumn{2}{|c|}{ Model Fit }& SGLM & SGLMM & SGLM & SGLMM &SGLM & SGLMM \\ \hline 
Linear Component & Data Set &&&&&&\\ \cline{1-2}
Intercept & 1 & 0.3859 & 0.3859 & 0.4048 & 0.4048 & 0.2789 & 0.2789 \\ 
& 2 & 0.4411 & 0.4411 & 0.3861 & 0.3861 & 0.5119 & 0.5154 \\ 
& 3 & 0.3432 & 0.3432 & 0.2774 & 0.2774 & 0.4595 & 0.4561 \\ 
\hline
Simple-1 & 1 & 0.3446 & 0.348 & 0.3082 & 0.3151 & 0.2189 & 0.2256 \\ 
& 2 & 0.3255 & 0.3154 & 0.3204 & 0.3204 & 0.2483 & 0.2483 \\ 
& 3 & 0.3643 & 0.3608 & 0.3463 & 0.3463 & 0.2337 & 0.2337 \\ 
\hline
Simple-2 & 1 & 0.3525 & 0.3559 & 0.3163 & 0.3129 & 0.1862 & 0.1793 \\ 
& 2 & 0.3127 & 0.3127 & 0.2603 & 0.2603 & 0.2407 & 0.2407 \\ 
& 3 & 0.3119 & 0.3017 & 0.3456 & 0.3456 & 0.2114 & 0.2114 \\ 
\hline
Multiple & 1 & 0.3241 & 0.3207 & 0.2619 & 0.2619 & 0.1875 & 0.1809 \\ 
& 2 & 0.3182 & 0.3252 & 0.2586 & 0.2586 & 0.2125 & 0.2125 \\ 
& 3 & 0.2688 & 0.276 & 0.2907 & 0.2907 & 0.2053 & 0.1987 \\ 
\hline
Confounded & 1 & 0.2601 &  0.2601 & 0.2712 & 0.2712  & 0.1507 &0.1507 \\ 
& 2 & 0.2226 & 0.2295 & 0.3223 & 0.3156 & 0.1313 & 0.1313\\ 
& 3 &  0.2616 & 0.2682 & 0.2721 & 0.2721 & 0.1729 & 0.1661 \\ 
\hline
\end{tabular}

\begin{tabular}{|c|c|cc|cc|cc|}
\hline
\multicolumn{8}{|l|}{ \textbf{ Test }}\\ \hline
\multicolumn{2}{|c|}{ $\kappa$ }& \multicolumn{2}{c|}{0.25} & \multicolumn{2}{c|}{0.5} & \multicolumn{2}{c|}{1}\\
 \multicolumn{2}{|c|}{ Model Fit }& SGLM & SGLMM & SGLM & SGLMM &SGLM & SGLMM \\ \hline 
Linear Component & Data Set &&&&&&\\ \cline{1-2}
Intercept & 1 & 0.4314 & 0.4314 & 0.4054 & 0.4324 & 0.2925 & 0.3302 \\ 
& 2 & 0.3981 & 0.3981 & 0.3711 & 0.3711 & 0.3084 & 0.2991 \\ 
& 3 & 0.3196 & 0.3196 & 0.2222 & 0.2407 & 0.2981 & 0.2981 \\ 
\hline
Simple-1 & 1 & 0.4712 & 0.4327 & 0.3704 & 0.3611 & 0.1553 & 0.1553 \\ 
& 2 & 0.3137 & 0.3235 & 0.3707 & 0.3621 & 0.1842 & 0.1842 \\ 
& 3 & 0.422 & 0.422 & 0.265 & 0.265 & 0.1101 & 0.1284 \\ 
\hline
Simple-2 & 1 & 0.2952 & 0.2952 & 0.3019 & 0.3208 & 0.1545 & 0.1727 \\ 
& 2 & 0.2936 & 0.3028 & 0.25 & 0.2593 & 0.1905 & 0.1905 \\ 
& 3 & 0.2857 & 0.2857 & 0.2745 & 0.2745 & 0.1667 & 0.1667 \\ 
\hline
Multiple & 1 & 0.3273 & 0.3182 & 0.2547 & 0.2642 & 0.1771 & 0.1875 \\ 
& 2 & 0.307 & 0.3333 & 0.2727 & 0.2818 & 0.1416 & 0.1327 \\ 
& 3 & 0.3802 & 0.3967 & 0.3604 & 0.3423 & 0.1122 & 0.1122 \\ 
\hline
Confounded & 1 & 0.3077 & 0.3269 & 0.3048 &  0.3143 &  0.2222 &0.2500 \\ 
& 2 & 0.3056 &0.3148 & 0.1717 & 0.1919 & 0.1262 &0.1359 \\ 
& 3 & 0.3367 & 0.3265 &0.2453 &  0.2453 & 0.2000 &0.1905 \\ 
\hline
\end{tabular}


\end{appendix}

\end{document}